\documentclass[conference]{IEEEtran}
\IEEEoverridecommandlockouts
\usepackage{cite}
\usepackage{amsmath,amssymb,amsfonts}
\usepackage{algorithmic}
\usepackage{graphicx}
\usepackage{textcomp}
\usepackage{xcolor}
\usepackage{flushend}
\usepackage{setspace}
\usepackage{xspace}
\usepackage{color}
\usepackage{colortbl, multirow}

\usepackage{hyperref}
\usepackage{url}
\usepackage{ascmac}

\usepackage{booktabs} 
\usepackage{subcaption} 
\def\BibTeX{{\rm B\kern-.05em{\sc i\kern-.025em b}\kern-.08em
    T\kern-.1667em\lower.7ex\hbox{E}\kern-.125emX}}
\begin{document}

\newcommand{\toolname}{ReOMP\xspace}
\newcommand{\mona}[1]{$\tt{#1}$}
\newcommand{\catalyst}{HOKUSAI BigWaterfall2\xspace} 

\newcommand{\mutexlockunlock}{\mona{mutex\_lock/unlock}\xspace}
\newcommand{\mutexlock}{\mona{mutex\_lock}\xspace}
\newcommand{\mutexunlock}{\mona{mutex\_unlock}\xspace}
\newcommand{\nexttid}{\mona{next\_tid}\xspace}
\newcommand{\gclock}{\mona{global\_clock}\xspace}
\newcommand{\nextclock}{\mona{next\_clock}\xspace}

\newcommand{\gatein}{\mona{gate\_in}\xspace}
\newcommand{\gateout}{\mona{gate\_out}\xspace}
\newcommand{\gateinout}{\mona{gate\_in/out}\xspace}

\newcommand{\kmpccritical}{\mona{\_\_kmpc\_critical}\xspace}
\newcommand{\kmpcendcritical}{\mona{\_\_kmpc\_end\_critical}\xspace}
\newcommand{\kmpcreduction}{\mona{\_\_kmpc\_reduction}\xspace}
\newcommand{\kmpcendreduction}{\mona{\_\_kmpc\_end\_reduction}\xspace}

\newcommand{\localsum}{\mona{local\_sum}\xspace}
\newcommand{\gsum}{\mona{sum}\xspace}

\newcommand{\kento}[1]{{\color{violet}[Kento:~#1]}}

\newtheorem{condition}{Condition}
\definecolor{darkgray}{gray}{0.7}
\definecolor{lightgray}{gray}{0.9}
\definecolor{floralwhite}{RGB}{255,250,240}

\title{
Distributed Order Recording Techniques for \\
\centering Efficient Record-and-Replay of \\
\centering Multi-threaded Programs
}

\author{%
  \IEEEauthorblockN{Xiang Fu$^1$, Shiman Meng$^1$, Weiping Zhang$^1$, Luanzheng Guo$^2$, Kento Sato$^3$, \\Dong H. Ahn$^4$, Ignacio Laguna$^5$, Gregory L. Lee$^5$, Martin Schulz$^6$}
  \IEEEauthorblockA{$^1$Nanchang Hangkong University, $^2$Pacific Northwest National Laboratory, $^3$RIKEN R-CCS, \\$^4$NVIDIA, $^5$Lawrence Livermore National Laboratory, $^6$Technical University of Munich\\
  \textit{\{fuxiang, smeng, wzhang\}@nchu.edu.cn}; 
  \textit{lenny.guo@pnnl.gov}; 
  \textit{kento.sato@riken.jp}; \\
  \textit{donga@nvidia.com};
  \textit{\{lagunaperalt1,lee218\}@llnl.gov};
  \textit{schulzm@in.tum.de}
  }
}

\maketitle

\begin{abstract}

After all these years and all these other shared memory programming frameworks, OpenMP is still the most popular one. However, its greater levels of non-deterministic execution makes debugging and testing more challenging. The ability to record and deterministically replay the
program execution is key to address this challenge.
However, scalably replaying OpenMP programs
is still an unresolved problem.
%
In this paper, we propose two novel techniques that use Distributed Clock (DC) and Distributed Epoch (DE) recording schemes to eliminate excessive thread synchronization for OpenMP record and replay.
%
%
%
Our evaluation on representative HPC applications with \toolname, which we used to realize DC and DE recording, shows that our approach is 2-5x more efficient than  traditional approaches that synchronize on every shared-memory access.
Furthermore, we demonstrate that our approach can be easily combined with MPI-level replay tools to replay non-trivial MPI+OpenMP applications. 
We achieve this by integrating \toolname into ReMPI, an existing scalable MPI record-and-replay tool, with only a small MPI-scale-independent runtime overhead.
%
\end{abstract}


\maketitle

\section{Introduction}
High-Performance Computing (HPC) has accelerated scientific discoveries
with faster and larger-scale simulations on continuously advancing hardware,
driven by the development of multi-core and many-core processors.
To fully utilize these hardware components and efficiently run scientific
simulations at large scale, application developers write
\emph{hybrid parallel applications}, which involves 
shared-memory and distributed-memory programming~\cite{Yang:2011,
Rabenseifner:2009, Wolf:2003, Mininni:2011}.
Since the largest supercomputers have been on the order of 
tens of millions of
compute cores~\cite{top500, sierra, summit} 
, hybrid parallel programming~(e.g., MPI~\cite{MPI}+OpenMP~\cite{OpenMP}) has become necessary to
realize the full potential of the available parallelism.

Hybrid parallel programming models introduce significant complexity to
applications~\cite{Wolf:2003, Drosinos:2004}.
Common programming patterns involve asynchronous
algorithms, where each thread and process asynchronously performs its own local computation and
then periodically exchanges or reduces the numerical results into shared
variables with other threads and processes via intra-process communication (e.g.,
shared-memory access) or inter-process communication (e.g., message
passing).
These asynchronous algorithms can introduce non-determinism, i.e., applications may behave
differently and/or output different numerical results in each
execution~\cite{Archer:2016, DOpenMP:2011, NINJA:2017,  MPIWiz:Xue:2009}.
Non-determinism makes debugging and testing these applications difficult
since programmers may not be able to reproduce buggy runs or target control
flows that they want to diagnose and test~(More details are discussed in Section~\ref{ssec:motivation:debugging}).

Record-and-replay is a promising approach to facilitate debugging and testing
non-deterministic applications that use message
passing~\cite{CDC:2015, MPIWiz:Xue:2009} and
shared memory~\cite{Chimera:2012, iDNA:2006}.
For message-passing applications, this approach records the order of received messages and then
uses these recordings to replay the exact same order of received messages in
subsequent replay runs. 
For the shared-memory case, this approach reproduces shared-memory 
access behaviors by recording and replaying the order that threads write to shared memory.


When recording and replaying the execution of hybrid parallel applications,
one must carefully consider how to reduce overheads.
While recording and replaying message passing is inexpensive~(e.g., runtime
overhead is typically about 10\% to 40\%~\cite{CDC:2015, Kranzlmuller:2001}),
recording and replaying shared-memory accesses is expensive~(e.g., runtime overhead ranges from
one to two orders of magnitude~\cite{Chimera:2012, PinPlay:2010}).
Reducing the overhead of shared-memory record-and-replay is critical for efficient
record-and-replay of hybrid parallel applications in practice.
In order to record the same shared-memory access order as observed and correctly replay as 
recorded,
traditional approaches serialize a series of operations for record-and-reply, which leads to 
excessive between-threads synchronizations~\cite{Chimera:2012, iDNA:2006}.
This \emph{excessive thread serialization and synchronization} for record-and-replay become the 
primary overhead of the record-and-replay approach for shared memory access since 
record-and-replay 
involves I/O operations to a file system (More details in Section~\ref{ssec:motivation:challenge}) .

In this paper, we propose two novel scalable \emph{distributed clock} (DC) and \emph{distributed epoch} (DE) \emph{recording} techniques. DC records
 logical clocks instead of thread IDs. Based on DC, we further optimize the method to improve the replay efficiency, which is DE. DE effectively reduces the serialization and synchronization requirements inherent in traditional recording methodologies through meticulous analysis and leveraging specific parallel conditions (outlined in detail in Condition~\ref{the:SMA:the1}). We observe that under the fulfillment of two parallel key conditions we induce, the execution order of multiple store or load instructions can be interchanged without affecting the final outcome of the program, which implies that these operations can be executed in parallel. We can use this for replay efficiency improvements.

Based on this observation, our DE technique allows each thread to independently record epochs in a distributed manner. Here, an epoch represents a logical time period. Each load or store instruction meeting the parallel conditions is assigned to its corresponding epoch, and instructions with the same epoch can be executed in parallel. By no longer mandating serialization or synchronization within epochs, our DE recording technique achieves significant optimizations in overhead compared to traditional approaches that overlook these parallel potentials. In the evaluation, we evaluated a total of 5 applications. In comparison to traditional techniques, with the use of 112 threads, the best-performing one is on a real HPC application HACC~\cite{HACC}, a sophisticated framework that employs particle-mesh techniques to simulate the evolutionary processes of mass within the universe, where DE recording can reduce overhead by 29\% in a record run and provide approximately 5.6 times speedup in a replay run.

To further validate the effectiveness and practicality of DE technique, we implement an OpenMP record-and-replay tool, \toolname, which incorporates our DE recording and combine it with an open-sourced MPI record-and-replay library, ReMPI~\cite{CDC:2015, ReMPI}. We apply ReMPI+ReOMP to two complex and non-deterministic OpenMP+MPI applications
. The results show that even when scaling from 24 OpenMP threads to 4800 threads involving varying node counts and process configurations, ReOMP maintains efficient record and replay capabilities, highlighting the potential of DE technique to reduce overhead and improve efficiency in large-scale parallel computing. Our techniques have good compatibility with existing multi-threaded record-and-replay tools, and can be run independently and in complementarity with existing record-and-replay tools. 
This paper makes the following 
contributions:
\begin{itemize}
  \item Two novel DC and DE recording techniques for efficient shared-memory record-and-replay;
  \item A quantitative performance evaluation and analysis of this implementation
  using synthetic OpenMP benchmarks and non-deterministic applications;
  \item A case study of ReOMP+ReMPI on two OpenMP+MPI applications. 
\end{itemize}

%

\section{Needs for Replaying Hybrid Models}
\label{sec:motivation}
\subsection{OpenMP Adds Additional Complexity}
\label{ssec:motivation:debugging}
  
When debugging, one common approach is to use
\mona{print} statements since this method is generic and easy to use.
Another approach is to use a debugger, for instance Totalview~\cite{TV:2024} or
DDT~\cite{DDT:2024} for parallel applications or GDB for serial applications.
While these approaches are effective in finding many classes of
bugs, they often do not suffice in finding bugs that arise
in the presence of non-determinism.
Bugs in non-deterministic applications may manifest themselves only under
particular control flows and thus may not always manifest themselves
during debug runs.
Even more, in the presence of additional print statements or when under control
of the debugger, the erroneous behavior may never get reproduced,
making debugging extremely challenging.

Consider a group of scientists in our organization who suffered from such
a non-deterministic bug: we recently assisted a production-application team
who had spent several months worth of effort, over the course of a year and a half,
finding and fixing a bug that only manifested once every 30 executions on average,
and only after a several hours of running. Due to the infrequent manifestation of the
bug, they eventually gave up on debugging even after spending significant amounts of
their time and computational resources.

Non-determinism hampers not only debugging, but also testing, particularly when it outputs different numerical results from one run to another. Many applications validate code modifications by comparing the numerical results to those produced from the previous code version. However, if the application itself is non-deterministic, it puts an enormous burden to the developers to determine whether the cause of any numerical difference is indeed due to the code modification. Overall, an ability to reproduce a particular control flow can significantly help developers to test non-deterministic applications.

When reproducing a particular numerical result, we must reproduce all
of the non-deterministic events which have affected the control flow
and the numerical results.
Consider another group of scientists in our organization, who
faced inconsistent numerical results from run to run
when they added OpenMP into a parallel application.
Although we applied ReMPI to try to reproduce the numerical results for them, replaying only
MPI message receives did not sufficiently reproduce the behavior
of the hybrid parallel application. Non-deterministic thread scheduling added another
level of non-determinism through intra-process (or inter-thread) communication
and synchronization,
i.e., shared memory accesses in reduction operations, critical sections, atomic operations, and benign data races.
Furthermore, we discovered that the thread scheduling even altered MPI message exchanges.

\subsection{Challenges for Shared-Memory Replay}
\label{ssec:motivation:challenge}
In our studies of hybrid applications, we have found that shared-memory
accesses occur much more frequently than message passing, thus reducing the
overhead of recording and replaying shared-memory accesses is of particular
importance. 
However, traditional record-and-replay approaches serialize a series of
thread operations for which leads to excessive slowdowns~\cite{Chimera:2012, iDNA:2006}.
Furthermore, since record-and-replay involves I/O operations to a file system,
the scalability of any record-and-replay tool is ultimately bounded by its file
system usage.
Therefore, relaxing thread serialization and more efficient I/O usage are critical
for recording and replaying threaded applications.
Our approaches, distributed clock recording (DC recording in Section~\ref{ssec:SMA:DC}) and distribute epoch recording (DE recording in Section~\ref{ssec:SMA:DE}), address these
critical technical challenges. 
\section{\toolname: Tool Flow}
\label{ssec:design:record-and-replay}

\begin{figure}[t]
	\centering
 	\includegraphics[width=7cm]{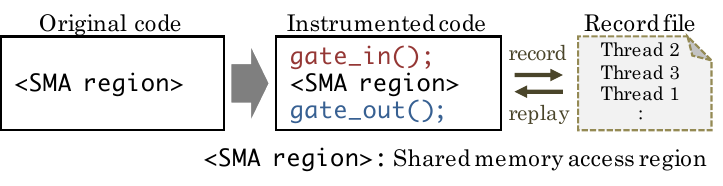}
 	\caption{The \gateinout functions}
 	\label{fig:design:gate_in_out}
\end{figure}

\begin{figure}[t]
	\centering
 	\includegraphics[width=8cm]{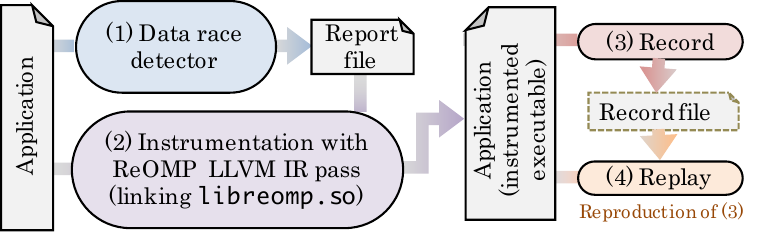}
 	\caption{Design overview}
 	\label{fig:design:overview}
\end{figure}

In shared-memory applications, non-determinism may occur when threads access
shared variables in a different order from a previous run.
We can reproduce a particular run by recording and replaying the order of 
shared variable accesses made by threads.
To accomplish this, we add instrumentation functions
(\gateinout) before and after a code region where shared-memory accesses occur
(Figure~\ref{fig:design:gate_in_out}).
The \gateinout functions are responsible for recording the order of threads
entering \emph{shared memory access regions} (SMA region in
Figure~\ref{fig:design:gate_in_out}) in a record run and for replaying the same order
in subsequent replay runs. The
shared-memory access region can be a single instruction or multiple
instructions.

Typically, in a multi-threaded application, critical sections~(e.g., \mutexlockunlock functions) and/or atomic
instructions are used when multiple threads may simultaneously access shared
variables.
We can easily identify these function calls and instructions, and then instrument them with our \gateinout functions.
\emph{Data races}, a condition where two or more threads concurrently access
the same memory location and at least one of these accesses is a write
instruction, are another source of simultaneous shared-memory accesses.
Unlike critical sections and atomic instructions, there is no
obvious indication of data races in source codes. 

To bridge this gap, we first apply a data-race detector called Tsan~\cite{Tsan} which integrated within clang to detect those races~(step (1) in Figure~\ref{fig:design:overview}). By compiling with \emph{-g} and \emph{-fsanitize=thread}, it detects data races in the application at runtime, capturing information such as function call stacks, absolute paths, line numbers, column numbers and etc., and generates a report file. 
Using function call information as input, we generated a unique hash value to create a data race instance.
These hash values will serve as the thread lock ID to control the thread access order to shared memory.
~\footnote{\toolname does not require the application developers to fix
the detected data races at this phase since the primary objective of use of \toolname is not \emph{fixing bugs} but \emph{replaying
bugs}. 
In the \toolname workflow, users are advised to fix data races that are
regarded as actual bugs.
However, even if the developers do not fix such bugs, it does not hamper the ability of
\toolname record-and-replay.
}

Next, we instrument all shared-memory-access regions in critical sections,
atomic instructions and data races with \gateinout~(step (2) in Figure~\ref{fig:design:overview}). 
We then run the instrumented executable to record the application's behavior (step (3) in Figure~\ref{fig:design:overview}). 
Finally, we replay the behavior based on the \emph{\toolname record file} for debugging and testing (step (4) in Figure~\ref{fig:design:overview}).

\section{Scalable Recording Techniques for Shared Memory Accesses}
\label{sec:SMA}
As explained in Section~\ref{sec:motivation}, efficient shared-memory
access recording is critical for reducing the overall overhead
of recording and replaying hybrid parallel applications.
In this section, we first describe a traditional approach before presenting
two novel techniques that can significantly improve upon it.
We choose to introduce these techniques gradually as each
later technique builds on the previous solution.

The traditional approach is referred as \emph{serialized thread ID recording}~(ST recording)
which records the order of \emph{thread~IDs} entering the \gateinout functions (in Section~\ref{ssec:SMA:ST}).
In contrast, our \emph{distributed clock recording}~(DC recording)
records the logical \emph{clocks} at which threads enter \gateinout~(in Section~\ref{ssec:SMA:DC}).
In Section~\ref{ssec:SMA:ST_vs_DC}, we compare ST and DC recording with respect
to the record-and-replay efficiency.
Finally, our \emph{distributed epoch recording}~(DE recording)
further improves upon them by recording
the logical \emph{epoch} during which threads can concurrently enter \gateinout
without breaking the correctness of replay~(in Section~\ref{ssec:SMA:DE}).


\subsection{Serialized Thread ID Recording (ST Recording)}
\label{ssec:SMA:ST}


\begin{figure}[t]
	\centering
 	\includegraphics[width=8cm]{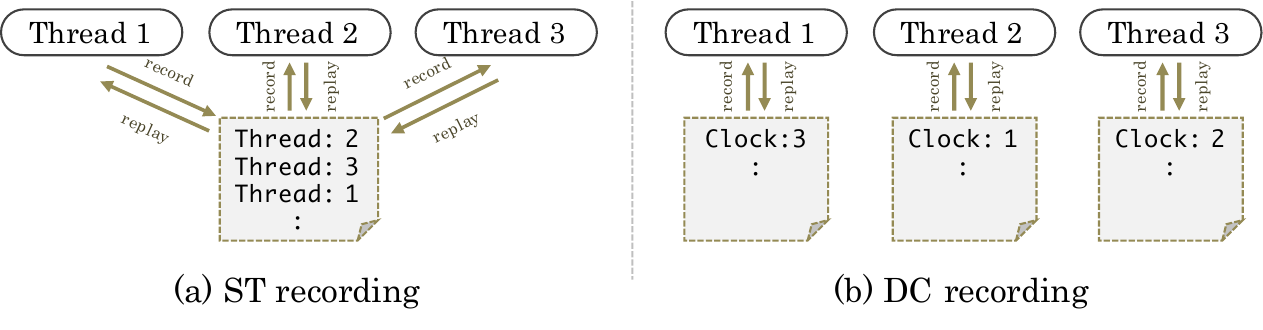}
 	\caption{ST and DC recording}
 	\label{fig:sma:st_dc_de}
\end{figure}

\begin{figure}[t]
	\centering
 	\includegraphics[width=8.5cm]{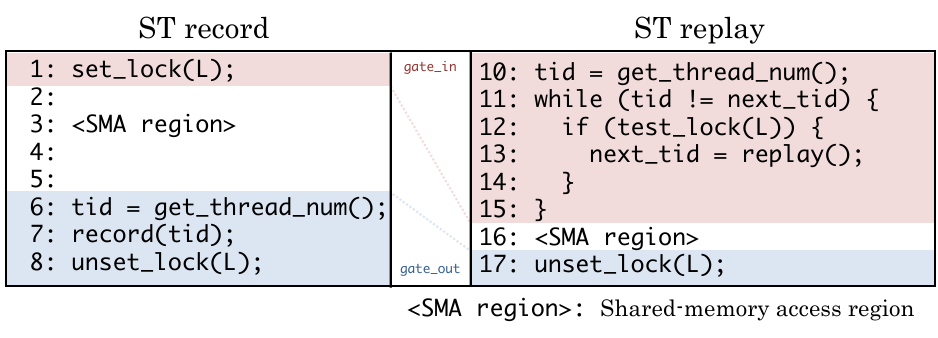}
 	\caption{ST record-and-replay}
 	\label{fig:sma:st_record_and_replay}
\end{figure}


For recording and replaying the order of concurrent shared-memory accesses,
a traditional approach is ST recording~(Figure~\ref{fig:sma:st_dc_de}-(a)).
This approach records thread IDs in the order that threads access shared
variables.

Figure~\ref{fig:sma:st_record_and_replay} shows the pseudo-code of ST recording.
In the record run, after a thread executes the shared-memory access region
(Line~3), the thread gets its thread ID~(Line~6) and records it to a file~(Line~7). Since we need to
record thread interleavings, we need to write these thread IDs to a single file.
In addition, depending on how the OS schedules threads, a thread~(e.g., Thread~X)
which already executed a shared-memory access region can be interleaved by
another thread~(e.g., Thread~Y). Then, if Thread~Y executes 
the shared-memory access region and then writes Thread~Y's ID to the file before
Thread~X writes its thread ID, the recorded order becomes Y$\rightarrow$X while the actual order is X$\rightarrow$Y.
Therefore, in order to ensure that order of recorded thread IDs is consistent
with the actual execution, we need to add
\mutexlockunlock to serialize the code from Line 1 to 8.

In replay runs for ST recording, each thread waits for its turn by polling a
variable, \nexttid~(Line~11). By updating \nexttid in the
recorded order of thread IDs, we can replay the order of threads accessing shared-memory
access regions. Since only one thread must read the record file to update \nexttid, 
each thread periodically tests the lock~(Line~12).
For example, if Thread~Z acquires the lock, then Thread~Z reads the next thread
ID and updates \nexttid~(Line~13). If \nexttid is ``W'', only Thread
W can exit the while-loop at Line~11 and then enter the shared-memory access region. Once the shared memory access
region is executed by Thread~W, Thread~W releases the lock so that other threads
can read the next thread ID from the record file in order to update \nexttid. Since we do not know which
thread needs to be replayed until we read the next thread ID value from the
file, all threads are candidates to read the file. Therefore, we need to have all
threads test the lock at Line~12. In ST recording, Line 1 (in record runs) and Line 10-15 (in replay runs)
correspond to \gatein, whereas Line 6-8 and Line 17
correspond to \gateout.

\subsection{Distributed Clock Recording (DC recording)}
\label{ssec:SMA:DC}


\begin{figure}[t]
	\centering
 	\includegraphics[width=8.5cm]{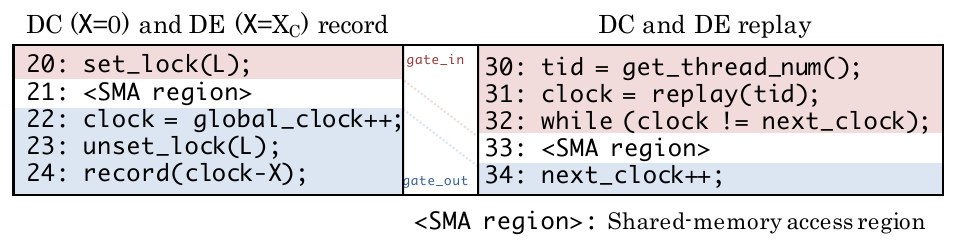}
 	\caption{DC (\mona{X}=0) and DE (\mona{X}=$X_C$) record-and-replay}
 	\label{fig:sma:dc_de_record_and_replay}
\end{figure}

With ST recording, we can correctly replay concurrent shared-memory accesses. 
However, there are a few disadvantages with respect to efficiency~(More details
in Section~\ref{ssec:SMA:ST_vs_DC}).
To resolve these issues, we propose a more efficient recording approach, DC
recording~(Figure~\ref{fig:sma:st_dc_de}-(b)). This approach records logical
clocks instead of thread IDs. In other words, DC recording records \emph{when}
threads enter each shared-memory access region, whereas ST recording records
\emph{which} threads enter each region.

Figure~\ref{fig:sma:dc_de_record_and_replay} shows the pseudo-code of DC recording where \mona{X} is always zero.
\mona{X} is used to compute epochs in DE recording in Section~\ref{ssec:SMA:DE}.
In record runs, after a thread executes the shared-memory access region~(Line~21), the thread
reads \gclock, increments~(Line~22) and records to its own file~(Line~24).
The \gclock variable is a global integer variable shared by
threads. By assigning a sequential clock number to each execution of a shared-memory access
region, each thread knows when to enter the region on successive replay
runs.
For the same reason as in ST recording, we must call \mutexlockunlock
functions in order to serialize code and assign correct sequential clock
numbers~(Line~20 and 23). With this approach, each thread can write clock
numbers to its own record file identified by its thread ID, \mona{tid}, 
as shown in Figure~\ref{fig:sma:st_dc_de}-(b).
Therefore, once a thread reads a current
\gclock value~(Line~22), the thread can immediately release the lock~(Line~23)
and write the value while other threads executes the region and get the
next \gclock value~(Line~24).

In replay runs, each thread reads the next clock value from
its own record file~(Line~31) and waits for its turn by polling a
variable, \nextclock~(Line~32). Once a thread completes a shared-memory access
region, the thread increments \nextclock by one so that the next
thread~(i.e., a thread having the next clock value, \nextclock+\mona{1}), can
enter the region.
By incrementing the \nextclock variable one by one, we can correctly replay shared-memory accesses as recorded.

In DC recording, Line~20 (in record runs) and Line 30-32 (in replay runs)
correspond to \gatein, whereas Line 22-24 and
Line~34 correspond to \gateout. Figure~\ref{fig:sma:st_dc_de}-(b)
shows DC recording when we record in order of Thread~$2\rightarrow3\rightarrow1$.
Although ST and DC approaches record the order in different ways, both approaches record the exact same order of thread acesses.
\subsection{ST versus DC Recording}
\label{ssec:SMA:ST_vs_DC}

\begin{figure*}[t]
\begin{tabular}{cc}
  \begin{minipage}{0.5\hsize}
  	\centering
 	\includegraphics[scale=0.45]{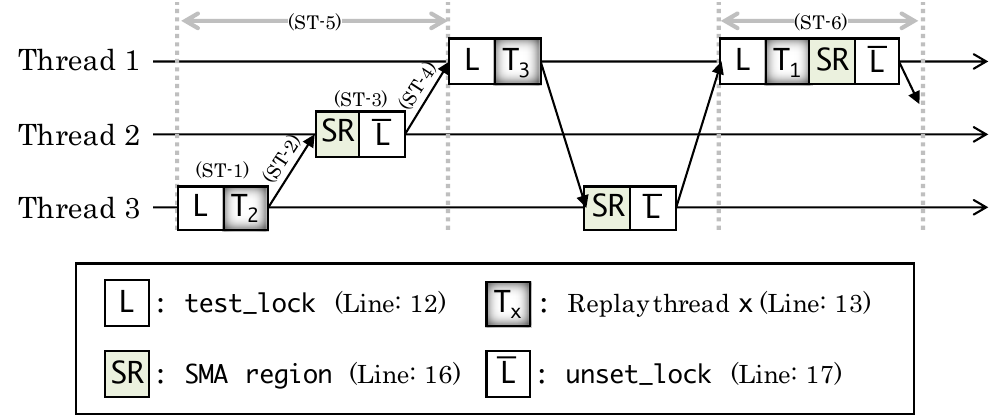}
 	\caption{ST replay}
 	\label{fig:sma:st_replay}
  \end{minipage}

  \begin{minipage}{0.5\hsize}
  	\centering
 	\includegraphics[scale=0.45]{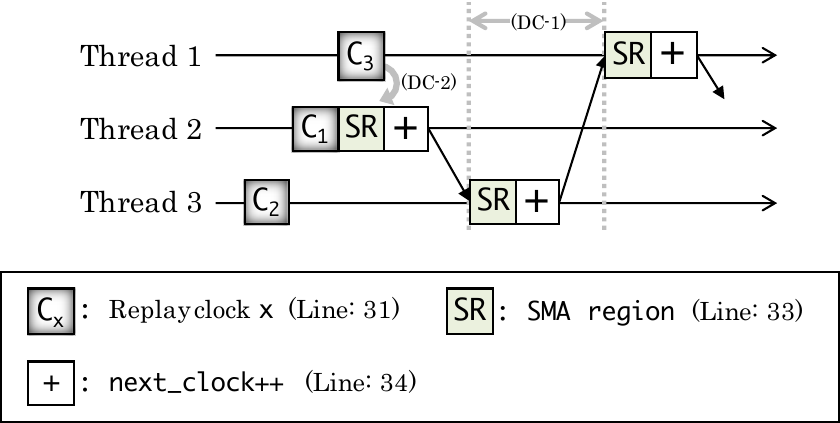}
 	\caption{DC replay}
 	\label{fig:sma:dc_replay}
  \end{minipage}

\end{tabular}
\end{figure*}

Although ST and DC record the exactly same order of thread accesses, DC
recording has several advantages. 
We explain the performance advantages of DC over ST recording
before explaining our final solution, DE recording.
\subsubsection{I/O throughput}
\label{sssec:SMA:io_throughput}
In ST recording, all I/O operations for record-and-replay must be serialized
since multiple threads concurrently write to and read from a single record file. 
In DC recording, all I/O operations are done in parallel since each
thread can write to and read from its own record file.
Therefore, multiple threads can issue I/O requests at the same
time. In general, parallel I/O can achieve higher I/O performance than
serialized I/O, thus DC recording is more efficient than ST recording with respect to
I/O throughput.

\subsubsection{Inter-thread communication}
\label{sssec:SMA:communication}
In the record runs, both ST and DC recording approaches need to call
\mutexlockunlock. Thus, there is no
difference between ST and DC recording with respect to
inter-thread communication in the record runs. However, ST recording issues
more inter-thread communication than DC recording in the replay runs.

Figure~\ref{fig:sma:st_replay} shows the example where order of thread IDs is
recorded as Thread 2$\rightarrow$3$\rightarrow$1~(The first three threads in
Figure~\ref{fig:sma:st_dc_de}-(a)).
In this example, Thread 3 acquires the lock and reads the next
thread ID to \mona{next\_tid}~(ST-1 in Figure~\ref{fig:sma:st_replay}). Since a
thread which reads the next thread ID (i.e., Thread 3) and the first replaying
thread ID (i.e., Thread 2) are different, the value of \nexttid in Thread 3
needs to be synchronized with \nexttid in Thread 2 via the global variable,
\mona{next\_tid}~(ST-2). 
Then, once Thread 2 completes the shared-memory access and releases the lock~(ST-3), the signal
is sent to the thread waiting for the lock, Thread 1 in this example~(ST-4).
Therefore, two inter-thread communications occur for replaying one execution of
a shared-memory access region~(ST-5). If a thread reading a file and a thread replayed happen to be the same, then only one inter-thread
communication occurs. However, as we increase the number of threads, it's less
likely that these two threads are identical.

In contrast, DC recording requires only one inter-thread communication
for every shared-memory access region~(DC-1 in Figure~\ref{fig:sma:dc_replay}).
In replay runs, the overhead of these inter-thread communications can become more
significant if threads running on different NUMA domains communicate (More details
in Section~\ref{ssec:evaluation:benchmarks}).

\subsubsection{I/O overlapping}
\label{sssec:SMA:io_overlapping}
The last advantage of DC recording is that each thread can overlap I/O with
other threads' I/O as well as execution of shared-memory access regions. 
In record runs of ST recording, each thread needs to write thread
IDs~(Line~7) between \mutexlock~(Line~1) and \mutexunlock~(Line~8),
whereas in DC recording, each thread can write clock values after
\mutexunlock~(Line~23 and 24) so that we can alleviate the I/O overhead by
overlapping with the I/O.
DC recording also can overlap I/O with other operations by
the same reason.
In the example in Figure~\ref{fig:sma:dc_replay}, Thread 1 reads clock 3
while Thread 2 executes the shared-memory access region~(DC-2).

\subsection{Distributed Epoch Recording (DE Recording)}
\label{ssec:SMA:DE}

\begin{table*}[t]
\tiny
\begin{tabular}{cc}
  \hspace{-0.5cm}
  
  \begin{minipage}[t]{0.2\hsize}
	  \caption{\footnotesize{Loads recorded}}
	  \label{tbl:sma:load_a}
	  \centering
	  \begin{tabular}{|c|c|c|}
	 \hline
	     T1 & T2 & T3 \\
		\hline
		Load X &  & \\
		& Load X  & \\
		&  & Load X \\
		\hline
			  \end{tabular}
  \end{minipage}
  \hspace{0.2cm}
  
  \begin{minipage}[t]{0.2\hsize}
	  \caption{\footnotesize{Loads replayed}}
	  \label{tbl:sma:load_b}
	  \centering
	  \begin{tabular}{|c|c|c|}
	  \hline
	     T1 & T2 & T3 \\	     
		\hline
		&  & Load X \\
		& Load X  & \\
		Load X &  & \\
		\hline
	  \end{tabular}
  \end{minipage}
  
  \hspace{0.3cm}

  \begin{minipage}[t]{0.3\hsize}
	  \caption{Stores recorded}
	  \label{tbl:sma:store_a}
	  \centering
	  \begin{tabular}{|c|c|c|}
	  \hline
	    T1 & T2 & T3 \\
		\hline
		Store 1\tiny{$\to$}X &  & \\
		& Store 2\tiny{$\rightarrow$}X  & \\
		&  & Store 3\tiny{$\rightarrow$}X \\
	    \hline
	  \end{tabular}
  \end{minipage}
	\hspace{-0.32cm}
  \begin{minipage}[t]{0.3\hsize}
	  \caption{Stores replayed}
	  \label{tbl:sma:store_b}
	  \centering
	  \begin{tabular}{|c|c|c|}
	  \hline
	  	T1 & T2 & T3 \\
		\hline
		& Store 2\tiny{$\rightarrow$}X  & \\
		Store 1\tiny{$\rightarrow$}X &  & \\
		&  & Store 3\tiny{$\rightarrow$}X \\
	    \hline
	  \end{tabular}
  \end{minipage}
\end{tabular}
\end{table*}

\begin{table}[t]
\caption{DC and DE recording}
\label{tbl:sma:DE}
\scriptsize
\centering
\begin{tabular}{|c|c|c|c|c|c|}
	\hline
    \rowcolor{darkgray}   & load or store	& Thread & (1) DC & (2) $X_C$ & (3) DE  \\
    \rowcolor{darkgray}   &                 &        & (clock)&           & ($epoch$)  \\
	\hline
	
	\cellcolor{lightgray} $x_0$  & \cellcolor{lightgray} Load X & \cellcolor{lightgray} T1 & 0 & 0 & 0 \\
	\hline 
	\cellcolor{lightgray} $x_1$  & \cellcolor{lightgray} Load X & \cellcolor{lightgray} T2 & 1 & 1 & 0 \\
	\hline
	\cellcolor{lightgray} $x_2$  & \cellcolor{lightgray} Load X & \cellcolor{lightgray} T3 & 2 & 2 & 0 \\
	\hline
	\cellcolor{lightgray} $x_3$  & \cellcolor{lightgray} Store 1$\rightarrow$X & \cellcolor{lightgray} T1 & 3 & 0 & 3 \\
	\hline
	\cellcolor{lightgray} $x_4$  & \cellcolor{lightgray} Store 2$\rightarrow$X & \cellcolor{lightgray} T2 & 4 & 1 & 3 \\
	\hline
	\cellcolor{lightgray} $x_5$  & \cellcolor{lightgray} Store 3$\rightarrow$X & \cellcolor{lightgray} T3 & 5 & 0 & 5 \\
	\hline
	\cellcolor{lightgray} $x_6$  & \cellcolor{lightgray} Load X & \cellcolor{lightgray} T1 & 6 & 0 & 6 \\
	\hline
\end{tabular}
\end{table}

\begin{table}[t]
\caption{Serialized(S) or paralleled/overlapped (P/O) operations in ST, DC and DE recording}
\label{tbl:sma:summary}
\scriptsize
\centering
\begin{tabular}{|c|c|c|c|}
	\hline
    \rowcolor{darkgray}   & ST & DC & DE \\
	\hline
	\cellcolor{lightgray} Getting thread ID or clock & S & S & S \\ 
	\hline 
	\cellcolor{lightgray} I/O for record-and-replay & S & P/O & P/O \\ 
	\hline
	\cellcolor{lightgray} Consecutive load and store instructions & S & S & P/O\\ 
	\hline
\end{tabular}

\end{table}

One drawback of ST and DC recording is that we still need to serialize every shared-memory access for correct replay.
DE recording, our final solution, relaxes this restriction without having to compromise the correctness of replay.
One key idea for DE recording is that we can still guarantee the replay correctness even if we change order of load and store instructions under Condition~\ref{the:SMA:the1}.
\begin{condition}
\label{the:SMA:the1}
Let $x_{i}$ be $i^{th}$ shared-memory access instruction, i.e., load or store instruction to the same memory address. Even if \toolname swaps order of $x_{i}$ and $x_{i+1}$ in replay runs, \toolname can still correctly replay shared-memory access when condition (i) or (ii) are held:
(i) $x_{i}$ and $x_{i+1}$ are both load instructions;
(ii) $x_{i}$, $x_{i+1}$ and $x_{i+2}$ are all store instructions.
\end{condition}
Tables~\ref{tbl:sma:load_a} and \ref{tbl:sma:load_b} are examples for condition (i).  
In Table~\ref{tbl:sma:load_a}, T1~(Thread 1) issues a load instruction to address $X$~(Load X) followed by load instructions from the same address $X$ by T2 and T3 in a record run. 
Since these three load instructions do not change memory state, T1, T2 and T3 can load the same value from address X, even if \toolname changes the order of these three load instructions in replay runs, e.g, T3$\rightarrow$T2$\rightarrow$T1 in Table~\ref{tbl:sma:load_b}.
 
Table~\ref{tbl:sma:store_a} and \ref{tbl:sma:store_b} are examples for condition (ii).
In Table~\ref{tbl:sma:store_a}, T1 stores value 1 to address $X$ (Store $1 \rightarrow X$) followed by store instructions to the same address $X$ by T2 and T3 in a record run.
Since the values, 1 and 2, by store instructions of T1 and T2 end up being overwritten by value 3 of the T3's store instruction, the value of X can be still replayed to be 3 after the completion of these store instructions even if \toolname swaps store instructions of T1 and T2 in the replay run as shown in Table~\ref{tbl:sma:store_b}.

This condition typically appears in applications which include intended data races (or benign data races~\cite{Boehm:2011}). In such applications, a group of threads (producers) periodically update variables with store instructions while the other group of threads (consumers) polls the variables until the variables are updated, i.e., busy-waiting or spinning techniques. Since busy-waiting and spinning techniques can remove expensive mutex lock and unlock operations, scientific applications tend to have this type of data races for user-level synchronization across threads.

To allow load and store instructions under Condition~\ref{the:SMA:the1} to be concurrently issued in replay runs, we record $epoch$ numbers instead of logical clocks in a record run. A epoch is a logical time period during which load or store instructions can be concurrently issued. For example, we assign the same epoch to the three instructions in Table~\ref{tbl:sma:load_a} and to the first two store instructions by T1 and T2 in Table~\ref{tbl:sma:store_a}.

Figure~\ref{fig:sma:dc_de_record_and_replay} shows the pseudo-code of DE recording where \mona{X} is $X_C$.
$X_C$ is used to compute epochs from clock, i.e., $epoch = clock - X_C$.
First, we get a logical clock in the same way as DC recording (Line 22) and then compute the epoch from $X_C$ before writing the clock (Line 24).
Table~\ref{tbl:sma:DE} shows an example how to compute $X_C$ and epoch.
For example, the second instruction, $x_2$, is a load instruction by T3.
Whenever a shared-memory access ($x_i$) occurs, \toolname computes the number of consecutive load or store instructions, $X_C$, in the past by looking up the memory-access history, i.e, $x_{i-1}, x_{i-2} \ldots$.
For example, at the $x_2$ load instruction, there were two consecutive load instructions at $x_0$ and $x_1$. Thus, $X_C$ becomes 2. Table~\ref{tbl:sma:DE}-(2) shows the values of $X_C$ in the example. 
Finally, we compute $clock - X_C (= epoch)$ and record the epoch number.
Since $x_4$ and $x_5$ are store instructions but $x_6$ is a load instruction, which violates (ii) in Condition~\ref{the:SMA:the1}, we can not swap $x_4$ and $x_5$.
To hold condition (ii), we set $X_C$ to 0 when a store instruction is followed by a load instruction so that the $x_6$ load instruction can always read the value 3 in replay runs.

Since shared-memory accesses can have the same clock number (i.e., epoch),
load or store instructions having the same clock can be concurrently executed while guaranteeing the correctness of the replay by using the same replay algorithm shown in Figure~\ref{fig:sma:dc_de_record_and_replay}.
In the example of Table~\ref{tbl:sma:DE}, \toolname replays in order of $\{x_0, x_1, x_2\} \rightarrow \{x_3, x_4\} \rightarrow \{x_5\} \rightarrow \{x_6\}$ where load or store instructions within the same brackets still can concurrently be executed with no serialization. 

To compute $X_C$, DE recording needs to keep the access history. We use a long-enough ring buffer so that the old access can automatically be discarded.  
In addition, since DE recording needs to look up the access history to compute $X_C$, the overhead of DE recording becomes slightly larger than that of DC recording (More details in Section~\ref{sec:evaluation}). However, DE recording allows concurrent load and store instructions under Condition~\ref{the:SMA:the1} during the replay runs. DE recording can improve replay performance. 
In practice, once we record an application run, we replay the run multiple times for debugging and testing the application, thus reducing replay overheads is more important than reducing record overhead.

Note that Condition~\ref{the:SMA:the1} is applied to only load and store instructions including atomic load and store and is not applied to other shared-memory accesses, e.g., reduction, critical section and atomic operations. In these other shared-memory access operations, the DE recording approach records in the same manner as DC recording.
As shown by synthetic benchmarks in Figure~\ref{fig:evaluation:omp_reduction}~to~\ref{fig:evaluation:data_race}, data races in which load and store instructions are concurrently executed lead to the largest overhead for record-and-replay of shared-memory accesses.
Therefore, DE recording is critical for reducing overall record-and-replay overhead in applications as shown by applications in Figure~\ref{fig:evaluation:amg_omp}~to~\ref{fig:evaluation:hpccg_omp}.
Table~\ref{tbl:sma:summary} summarizes what operations are serialized or parallelized (thereby, overlapped with other operations) in ST, DC and DE recording.
\section{Implementation}
To codify the above techniques, we develop a
record-and-replay tool to reproduce shared-memory accesses in applications that use
OpenMP.



As described in Section~\ref{ssec:design:record-and-replay}, shared-memory record-and-replay can be
achieved by instrumenting the \gateinout functions before and
after shared-memory regions. To instrument these functions, we use Clang/LLVM. The Clang compiler
provides an option to specify a shared library implementing an LLVM IR pass so that we can easily instrument the
\gateinout functions without modifying the application's source code. Then, we
link our record-and-replay shared library, \mona{libreomp.so}, which implements
the \gateinout functions, to applications in Figure~\ref{fig:design:overview}-(2).
We switch between record and replay modes with an environment variable. 

We now describe
how we instrument the most common shared-memory accesses
in OpenMP: critical, atomic and
reduction clauses.
For example in critical sections, Clang translates each critical clause into
\kmpccritical and \kmpcendcritical function calls, as
implemented in the LLVM OpenMP runtime library~\cite{LLVMOpenMP}. 
The \kmpccritical function is called before and the \kmpcendcritical function is called after a
critical section. By instrumenting \gatein before
\kmpccritical and \gateout after \kmpcendcritical,
we can correctly record and replay critical clauses. With this method, we can
also instrument other potential shared-memory accesses, such as the OpenMP reduction, the master and
the single clauses, by detecting a pair of called functions whose names respectively begin with
\mona{\_\_kmpc\_*} and
\mona{\_\_kmpc\_end\_*}. 
Clang translates each atomic clause into a corresponding atomic
instruction (e.g., \mona{atomicrmw} and \mona{cmpxchg}) so that we can easily
locate and instrument them with the \gateinout functions.

\section{Evaluation}
\label{sec:evaluation}

\begin{table}[t]
\caption{\catalyst Specification}
\label{tbl:evaluation:spec}
\centering
\scriptsize
\begin{tabular}{|c|l|}
	\hline
    \cellcolor{lightgray} Nodes 	& 312 nodes\\
	\hline
	\cellcolor{lightgray} CPU  		& Intel Xeon Max 9480 (1.9 GHz, 56 cores x 2 sockets) \\
	\hline
	\cellcolor{lightgray} Memory  	& 39TB(Memory Bandwidth: 3.26TB/s)\\
	\hline
	\cellcolor{lightgray} Interconnect  	& InfiniBand NDR(400GB/s between nodes)\\
	\hline
	
\end{tabular}
\end{table}

First, we show performance results in synthetic benchmarks in order to better understand
record-and-replay overhead of ST, DC and DE recording compared to the execution without record-and-replay in
Section~\ref{ssec:evaluation:benchmarks}. 
Then, we apply \toolname to five applications, AMG~\cite{AMG}, QuickSilver~\cite{QuickSilver, Richards:2017} , miniFE~\cite{CORAL_benchmarks}, HACC~\cite{HACC} and HPCCG~\cite{HPCCG} in Section~\ref{ssec:evaluation:proxy} in order to show the efficiency of DC and DE recording over ST recording. To demonstrate that our techniques can reduce overhead, the evaluations in these two sections were conducted with 2 to 112 threads running respectively. The distributed recording techniques we propose in this paper can be orthogonalized to existing record-and-replay tools, so we didn't compare with other existing tools.

Upon observation, it was found that if an MPI+OpenMP hybrid application, while utilizing multiple nodes and running multiple processes and threads, can't be replayed solely using ReMPI. Therefore, in Section~\ref{ssec:evaluation:case_study}, we show two cases of ReMPI+\toolname for recording and replaying two MPI+OpenMP applications: HACC, HPCCG. 

Our evaluation is conducted on a production system at our organization, \catalyst~\cite{HBW2}. 
Table~\ref{tbl:evaluation:spec} shows the specification of \catalyst. 
To enable OpenMP thread affinity, we set \mona{cpus-per-task} to \mona{OMP\_NUM\_THREADS}.
We store record files in a \mona{tmpfs} file system.

\subsection{Performance Study}
\label{ssec:evaluation:benchmarks}

\begin{figure}[t]
	\centering
 	\includegraphics[width=7cm]{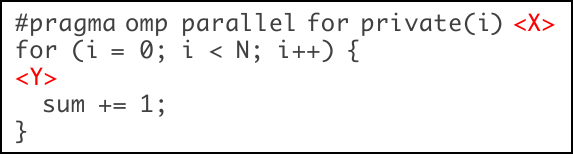}
 	\caption{Template of synthetic benchmarks}
 	\label{fig:evaluation:template}
\end{figure}

\begin{table}[t]
\caption{OpenMP clause for \mona{<X>} and \mona{<Y>} of
Figure~\ref{fig:evaluation:template}}
\label{tbl:evaluation:benchmarks}
\footnotesize
\centering
\begin{tabular}{|l|c|c|}
	\hline
    \rowcolor{darkgray}   	& \mona{<X>} & \mona{<Y>}  \\
	\hline
	\cellcolor{lightgray} omp\_reduction  	& \mona{reduction(+:sum)} & N/A \\
	\hline 
	\cellcolor{lightgray} omp\_critical  	& N/A & \mona{\#pragma\ omp\ critical} \\
	\hline
	\cellcolor{lightgray} omp\_atomic    	& N/A & \mona{\#pragma\ omp\ atomic} \\
	\hline
	\cellcolor{lightgray} data\_race  	    & N/A & N/A \\
	\hline
\end{tabular}
\end{table}

\makeatletter
\newcommand{\figcaption}[1]{\def\@captype{figure}\caption{#1}}
\newcommand{\tblcaption}[1]{\def\@captype{table}\caption{#1}}
\makeatother

\begin{table*}[t]
\begin{center}
	\begin{tabular}{cc}
		\begin{minipage}{5.5cm}
			\centering
 			\includegraphics[width=5.5cm]{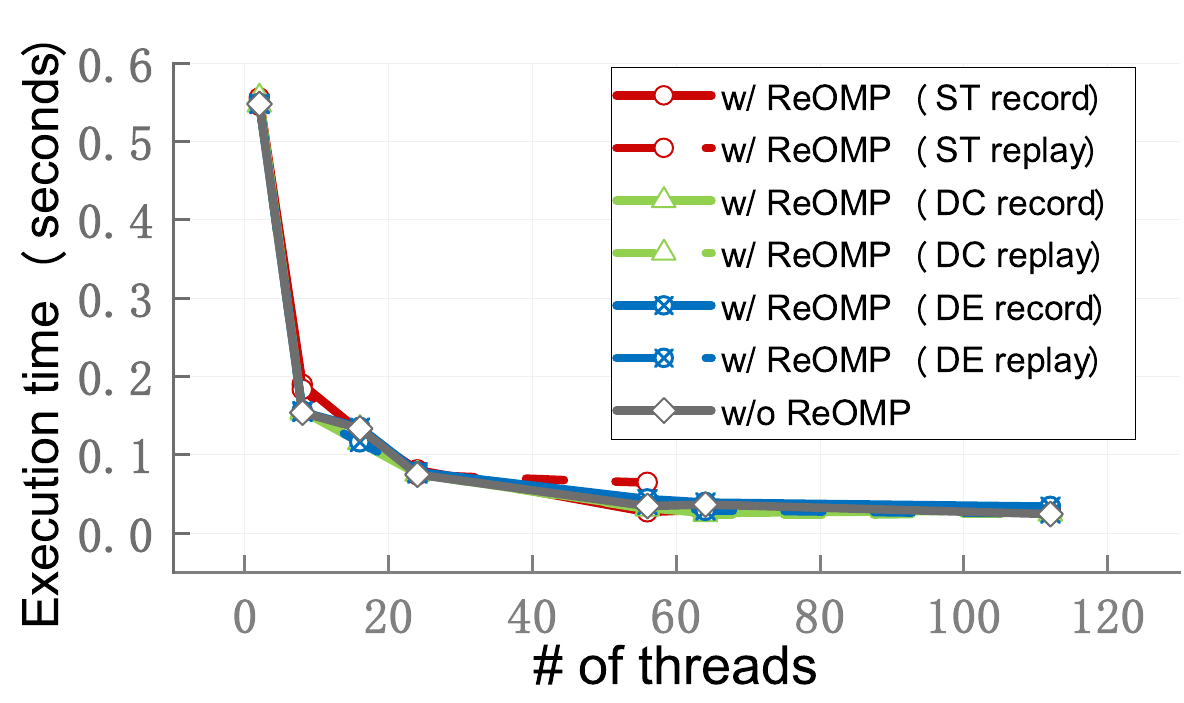}
 			\figcaption{Execution time of omp\_reduction}
 			\label{fig:evaluation:omp_reduction}
		\end{minipage}
		\hspace{0.1cm}
		\begin{minipage}{5.5cm}
			\centering
 			\includegraphics[width=5.5cm]{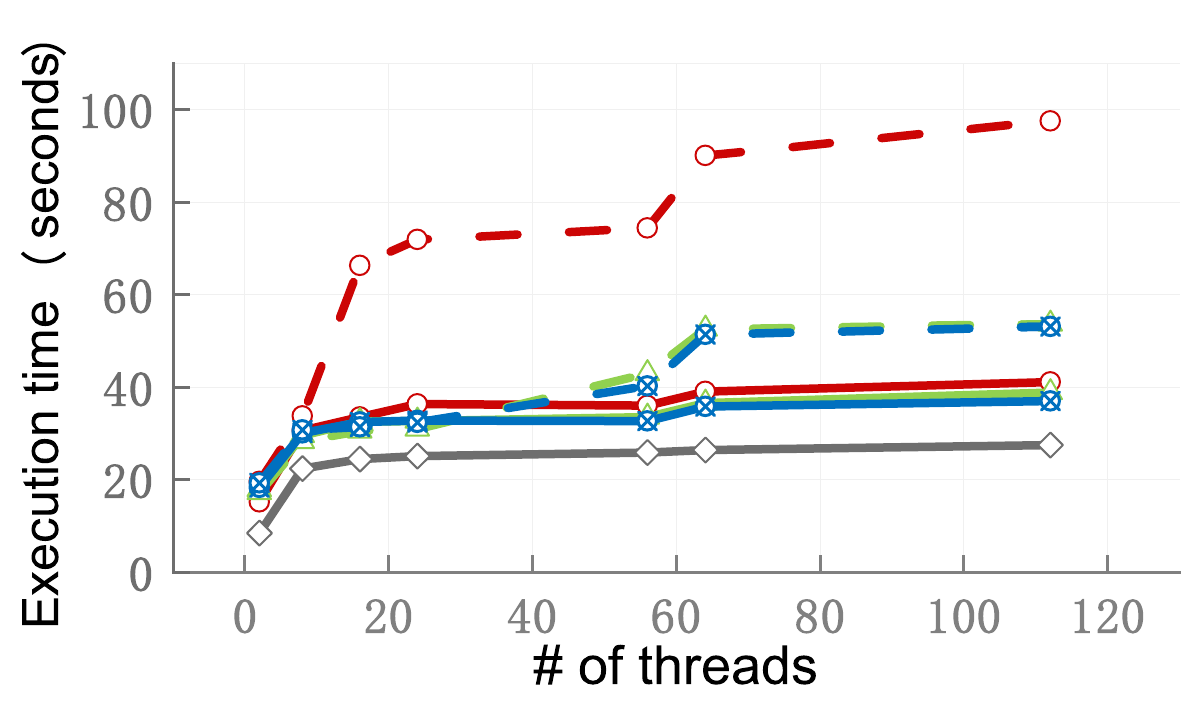}
 			\figcaption{Execution time of omp\_critical}
 			\label{fig:evaluation:omp_critical}
		\end{minipage}
		\hspace{0.1cm}
		\begin{minipage}{5.5cm}
			\centering
 			\includegraphics[width=5.5cm]{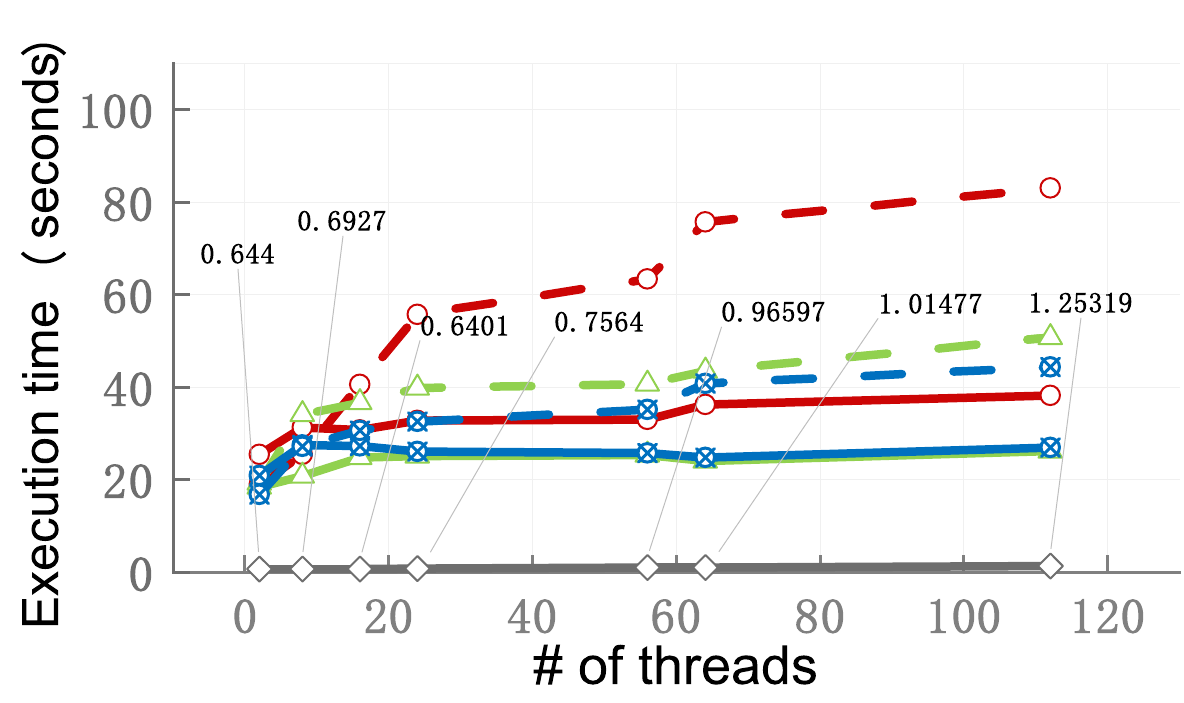}
 			\figcaption{Execution time of omp\_atomic}
 			\label{fig:evaluation:omp_atomic}
		\end{minipage}
		\\
		\begin{minipage}{5.5cm}
			\centering
 			\includegraphics[width=5.5cm]{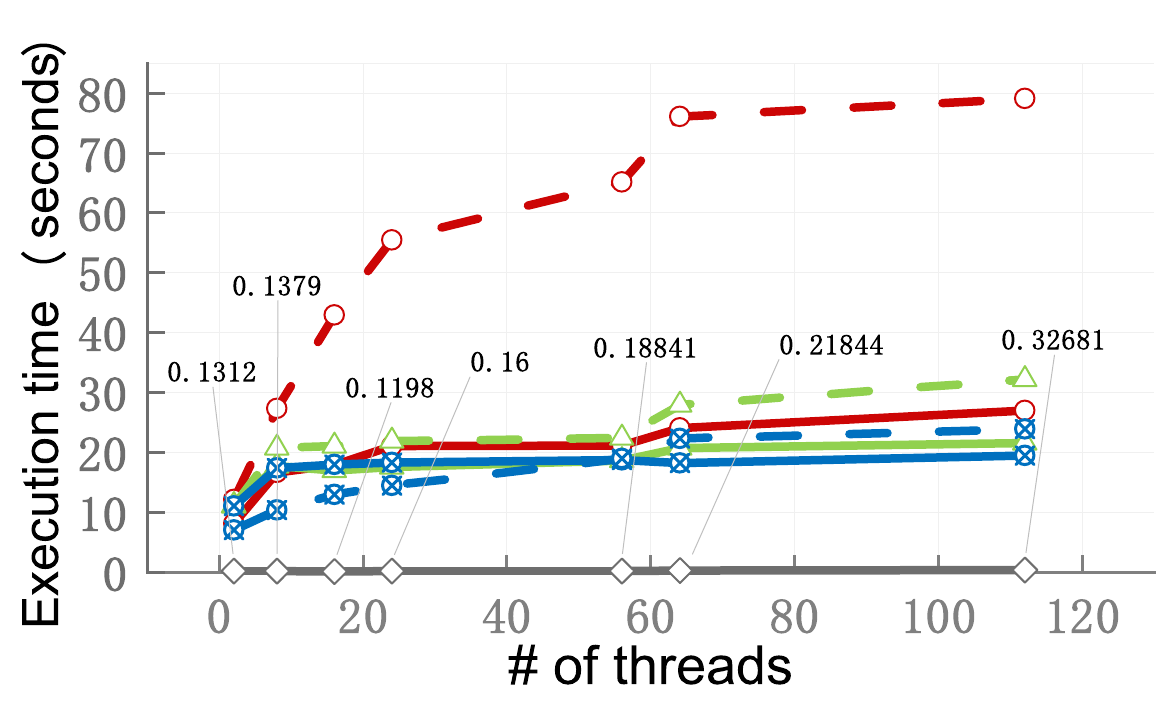}
 			\figcaption{Execution time of data\_race}
 			\label{fig:evaluation:data_race}
		\end{minipage}
		\hspace{0.1cm}
		\begin{minipage}{11.1cm}
			\def\@captype{table}
			\centering
			\footnotesize
			\begin{tabular}{|c|c|c|c|c|c|c|}
				\hline
				\rowcolor{darkgray}   	& \multicolumn{2}{c|}{ST} & \multicolumn{2}{c|}{DC}  & \multicolumn{2}{c|}{DE} \\
				\cline{2-7}
				\rowcolor{darkgray}   	& Record & Replay & Record & Replay & Record & Replay \\
				\hline
				\rowcolor{white}
				\cellcolor{lightgray} omp\_reduction  	& 1.23& 1.37& 1.2& 1.03& 1.37& 1.05\\
				\hline 
				\cellcolor{lightgray} omp\_critical  	& 1.49& 3.55& 1.41& 1.95& 1.34& 1.93\\ 
				\hline
				\cellcolor{lightgray} omp\_atomic    	& 30.54& 66.34& 20.15& 40.56& 21.51& 35.4\\
				\hline
				\cellcolor{lightgray} data\_race  	    & 82.46& 241.82& 65.86& 98.31& 59.57& 73.05\\
				\hline
			\end{tabular}
			\vspace{0.5cm}
    			\tblcaption{Relative execution times of ST, DC and DE approaches to the execution time w/o \toolname in 112 threads}
    			\label{tbl:evaluation:relative}
		\end{minipage}
		
	\end{tabular}
\end{center}
\end{table*}

First, we present our performance study of ST, DT and DE recording.
To establish a systematic understanding and measurement for record-and-replay overhead in each type of shared memory access, we use synthetic benchmarks. Figure~\ref{fig:evaluation:template} is a template of the
synthetic benchmarks. We implement four benchmarks, omp\_reduction,
omp\_critical, omp\_atomic and data\_race, by replacing \mona{<X>} and
\mona{<Y>} with OpenMP clauses as shown in
Table~\ref{tbl:evaluation:benchmarks}.
To avoid compiler optimization from removing the for-loop and statically assigning a constant value to \mona{sum} (i.e.,
\mona{sum=N}), we use a \mona{volatile} type for \mona{sum}.
For each benchmark, we iterate long enough to have execution time of the main
loop dominate the entire execution time.

\subsubsection{omp\_reduction}
\label{sssec:evaluation:reduction}

Figure~\ref{fig:evaluation:omp_reduction} shows the execution times of
omp\_reduction across the different number of threads. 
In the OpenMP reduction, each thread locally computes the
summation then reduces into the shared variable at the end. 
Thus, the overhead of the ST, DC and DE
record-and-replay methods is negligibly small in reduction because every thread
records and replays shared memory accesses only once at the end of the loop.

\subsubsection{omp\_critical}
\label{sssec:evaluation:critical}

Figure~\ref{fig:evaluation:omp_critical} shows the execution times of
omp\_critical. In the record runs, DC recording is faster than ST recording due
to the reasons described in Section~\ref{sssec:SMA:io_throughput} and
\ref{sssec:SMA:io_overlapping}. In addition to these overheads, the ST replay runs issue more inter-thread communication than the DC
replay runs as mentioned in Section~\ref{sssec:SMA:communication}. Therefore, 
the ST replay is much slower than DC replay.
In omp\_critical, DC and DE behave exactly the same way, thereby, 
the performance results are almost same. 

Our OpenMP affinity settings 
places threads onto the first socket until all cores in the socket are assigned and then onto the second socket next. The communication cost between threads across different NUMA domains is significantly higher compared to communication within the same NUMA domain. Consequently, once we run more threads than the number of cores of a single NUMA domain, overhead for record-and-replay becomes larger. 
For example, in HOKUSAI BigWaterfall2 with 56 cores, as depicted in Figure~\ref{fig:evaluation:omp_critical}, the overhead increases significantly when the number of threads increases from 56 to 64. Since ST replay runs involve more communication than DC/DE replay runs, the increase in overhead in ST replay runs is more pronounced than in DC/DE replay runs.

\subsubsection{omp\_atomic and dara\_race}

Figure~\ref{fig:evaluation:omp_atomic} and~\ref{fig:evaluation:data_race} show
the execution times of omp\_atomic and data\_race. With the same reasons in
omp\_critical, DC/DE recording is also more efficient than ST recording in
omp\_atomic and data\_race. 
In addition, DE can concurrently execute load and store instructions
under Condition~\ref{the:SMA:the1} in replay runs. Therefore, DE replays can run faster than DC replay.

Table~\ref{tbl:evaluation:relative} shows relative execution times of ST, DC and DE approaches to the one
without \toolname at 112 threads. The summation operation in
omp\_critical and omp\_atomic is originally serialized even without \toolname while
summation in data\_race is done concurrently. Therefore, the
overheads in omp\_critical and omp\_atomic are much smaller than the one in data\_race. 

\subsection{Non-deterministic Applications}
\label{ssec:evaluation:proxy}

\begin{table*}[t]
\begin{center}
	\begin{tabular}{ccc}
		\begin{minipage}{5.5cm}
			\centering
 			\includegraphics[width=5.5cm]{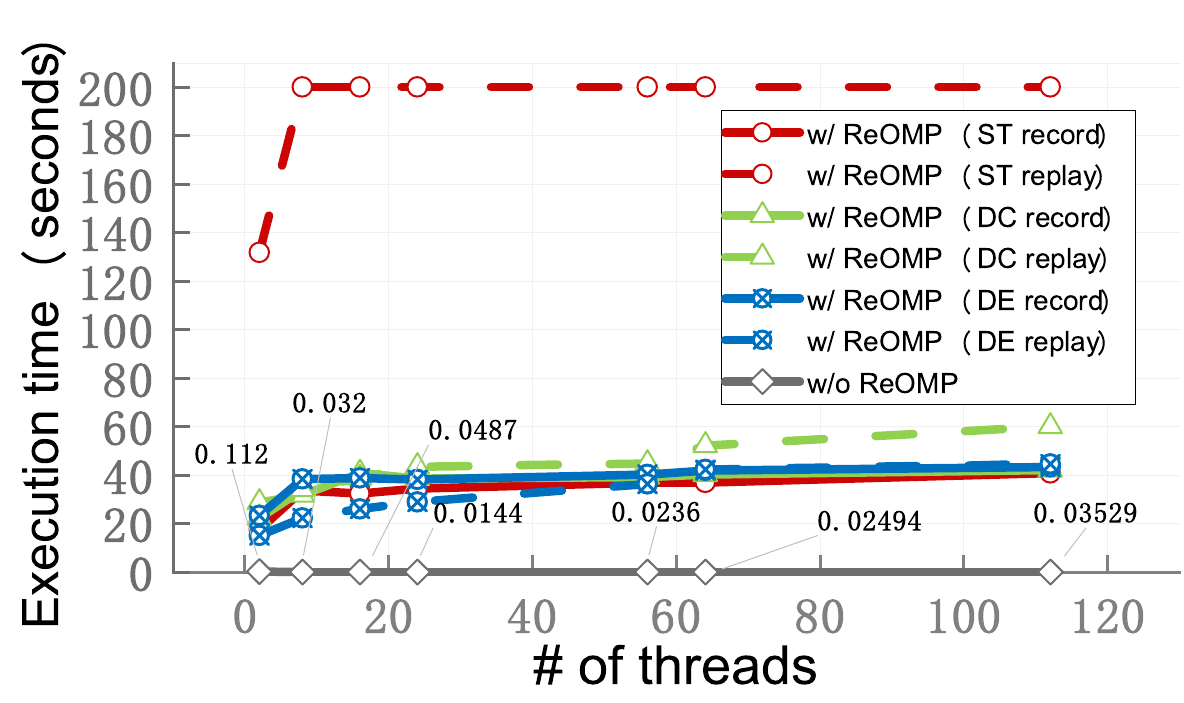}
 			\figcaption{Execution time of OpenMP AMG }
 			\label{fig:evaluation:amg_omp}
		\end{minipage}
		\hspace{0.1cm}
		\begin{minipage}{5.5cm}
			\centering
 			\includegraphics[width=5.5cm]{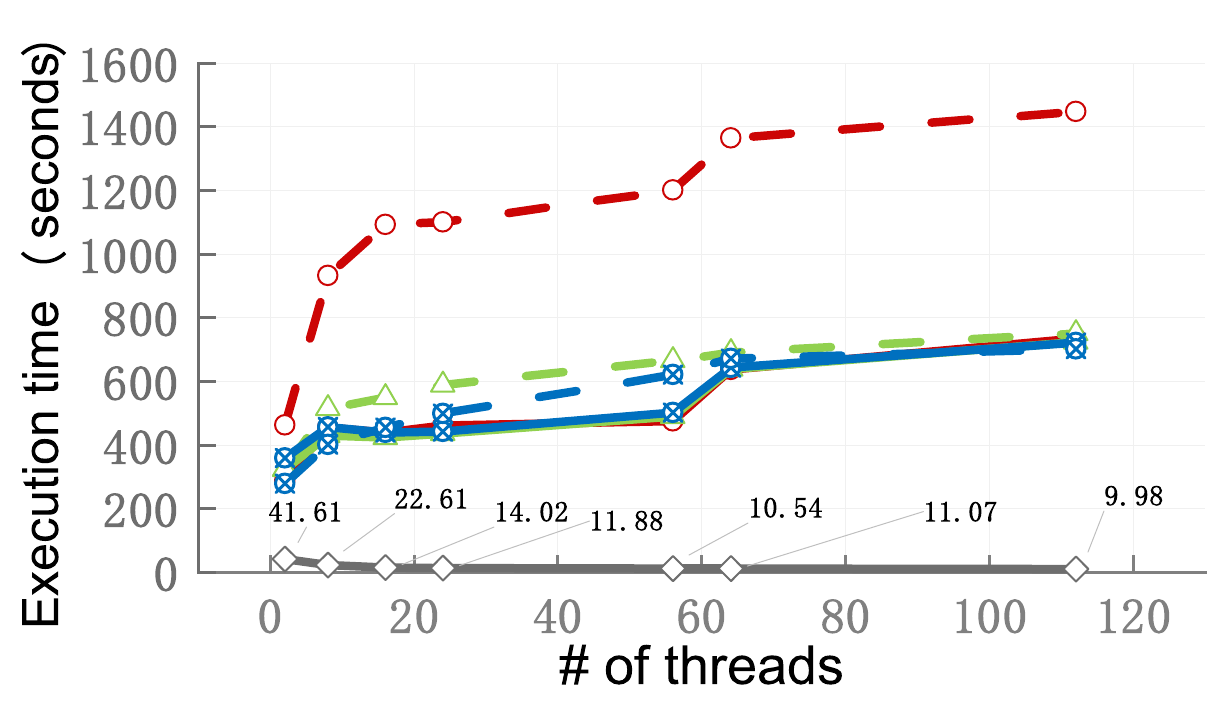}
 			\figcaption{Execution time of OpenMP QuickSilver}
 			\label{fig:evaluation:quicksilver_omp}
		\end{minipage}
		\hspace{0.1cm}
		\begin{minipage}{5.5cm}
			\centering
 			\includegraphics[width=5.5cm]{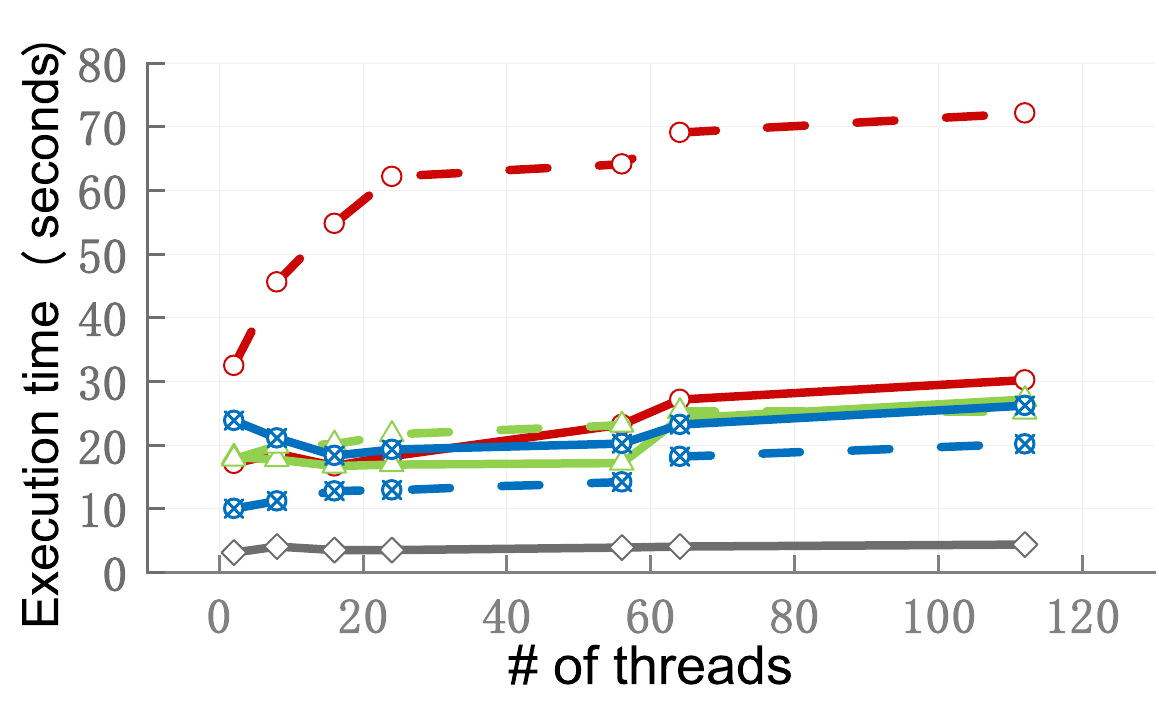}
 			\figcaption{Execution time of OpenMP miniFE}
 			\label{fig:evaluation:miniFE_omp}
		\end{minipage}
		\\
		\begin{minipage}{5.5cm}
			\centering
 			\includegraphics[width=5.5cm]{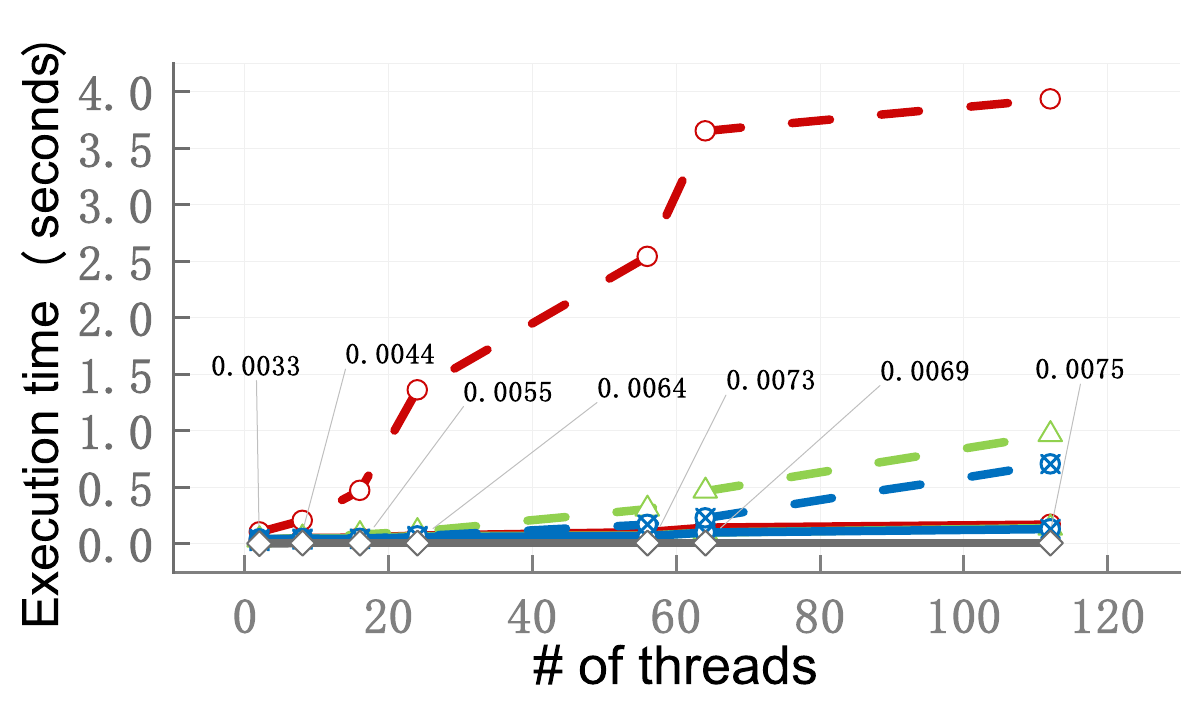}
 			\figcaption{Execution time of OpenMP HACC }
 			\label{fig:evaluation:hacc_omp}
		\end{minipage}
		\hspace{0.1cm}
		\begin{minipage}{5.5cm}
			\centering
 			\includegraphics[width=5.5cm]{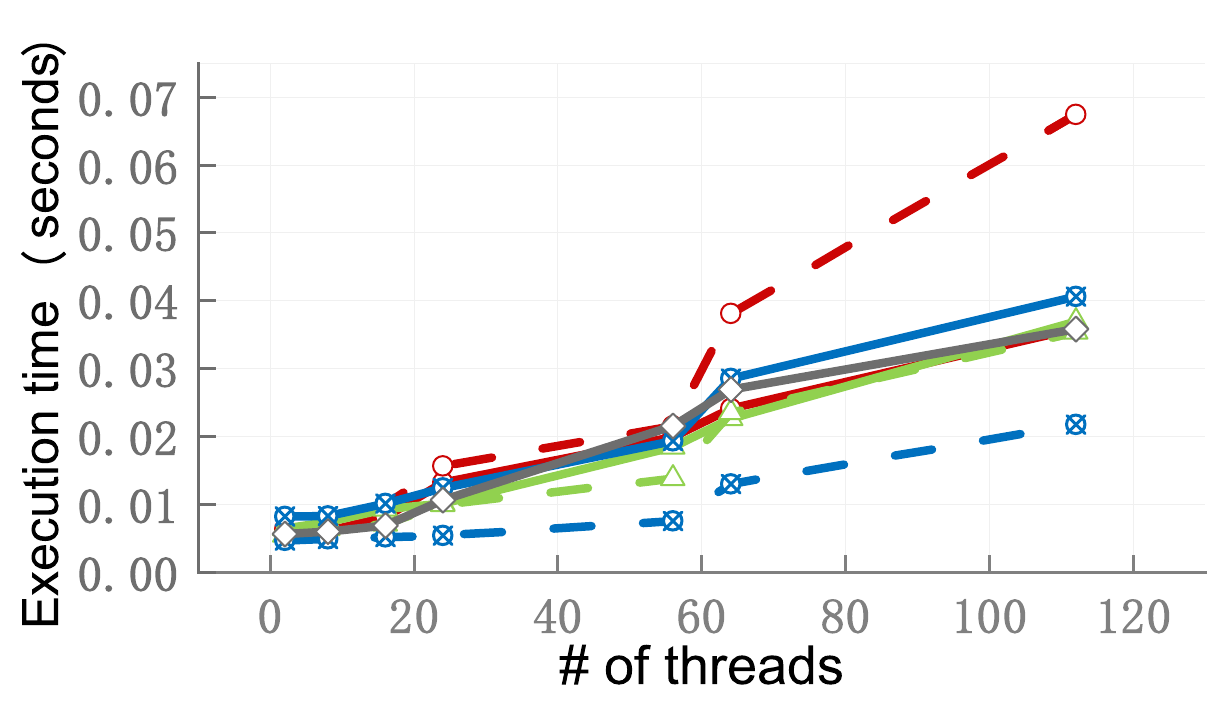}
 			\figcaption{Execution time of OpenMP HPCCG}
 			\label{fig:evaluation:hpccg_omp}
		\end{minipage}
		\hspace{0.1cm}
            \begin{minipage}{5.5cm}
			\centering
 			\includegraphics[width=5.5cm]{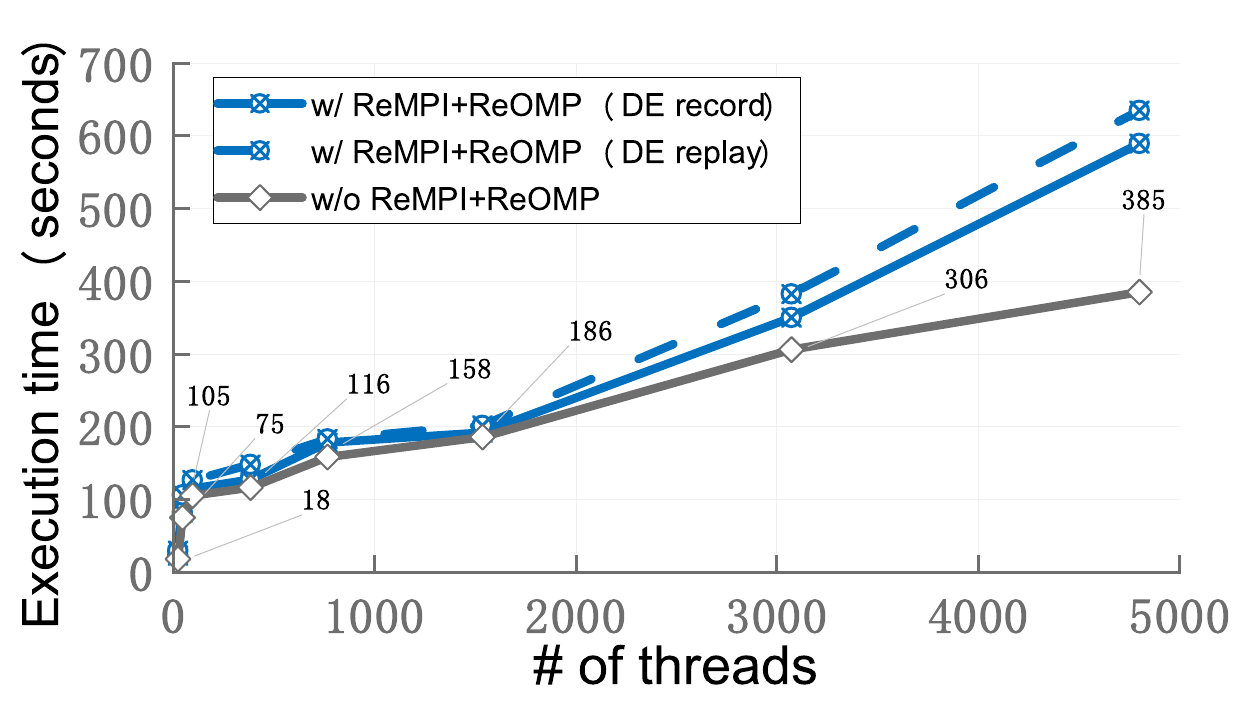}
 			\figcaption{Execution time of OpenMP+MPI HACC}
 			\label{fig:evaluation:mpi_omp_HACC}
		\end{minipage} 
            \\
		\begin{minipage}{5.5cm}
			\centering
 			\includegraphics[width=5.5cm]{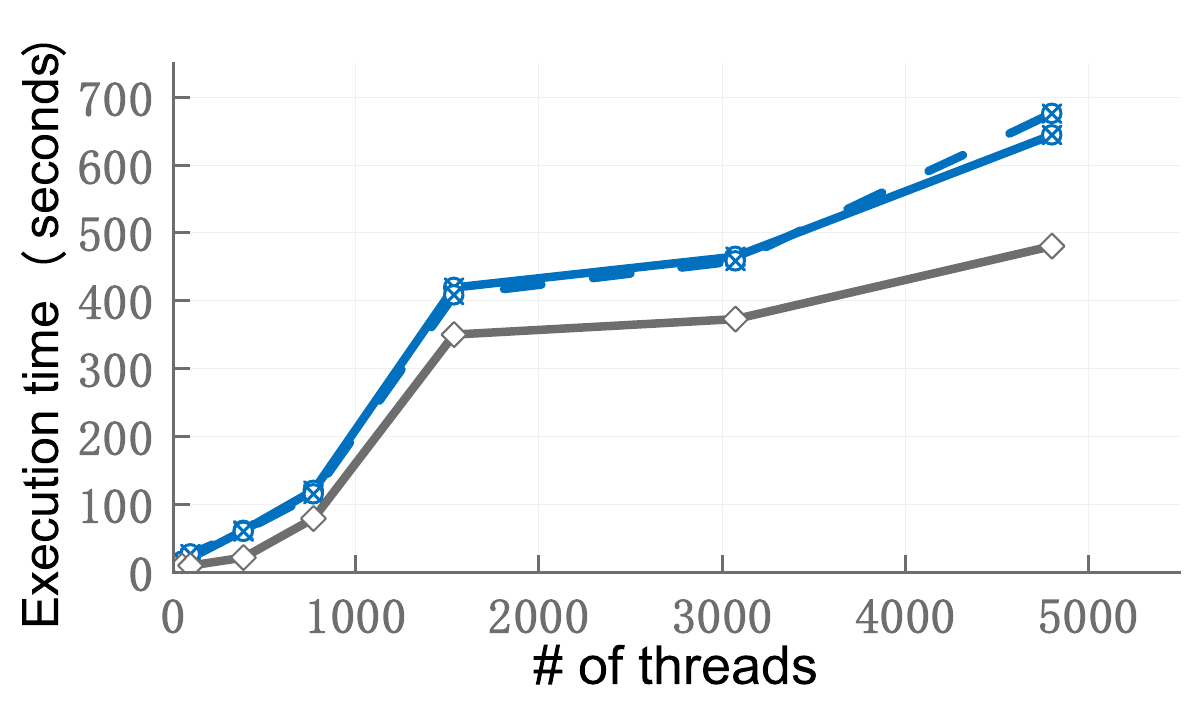}
 			\figcaption{Execution time of OpenMP+MPI HPCCG}
 			\label{fig:evaluation:mpi_omp_hpccg}
		\end{minipage}           
		\hspace{0.1cm}
		\begin{minipage}{5.5cm}
			\centering
 			\includegraphics[width=5.5cm]{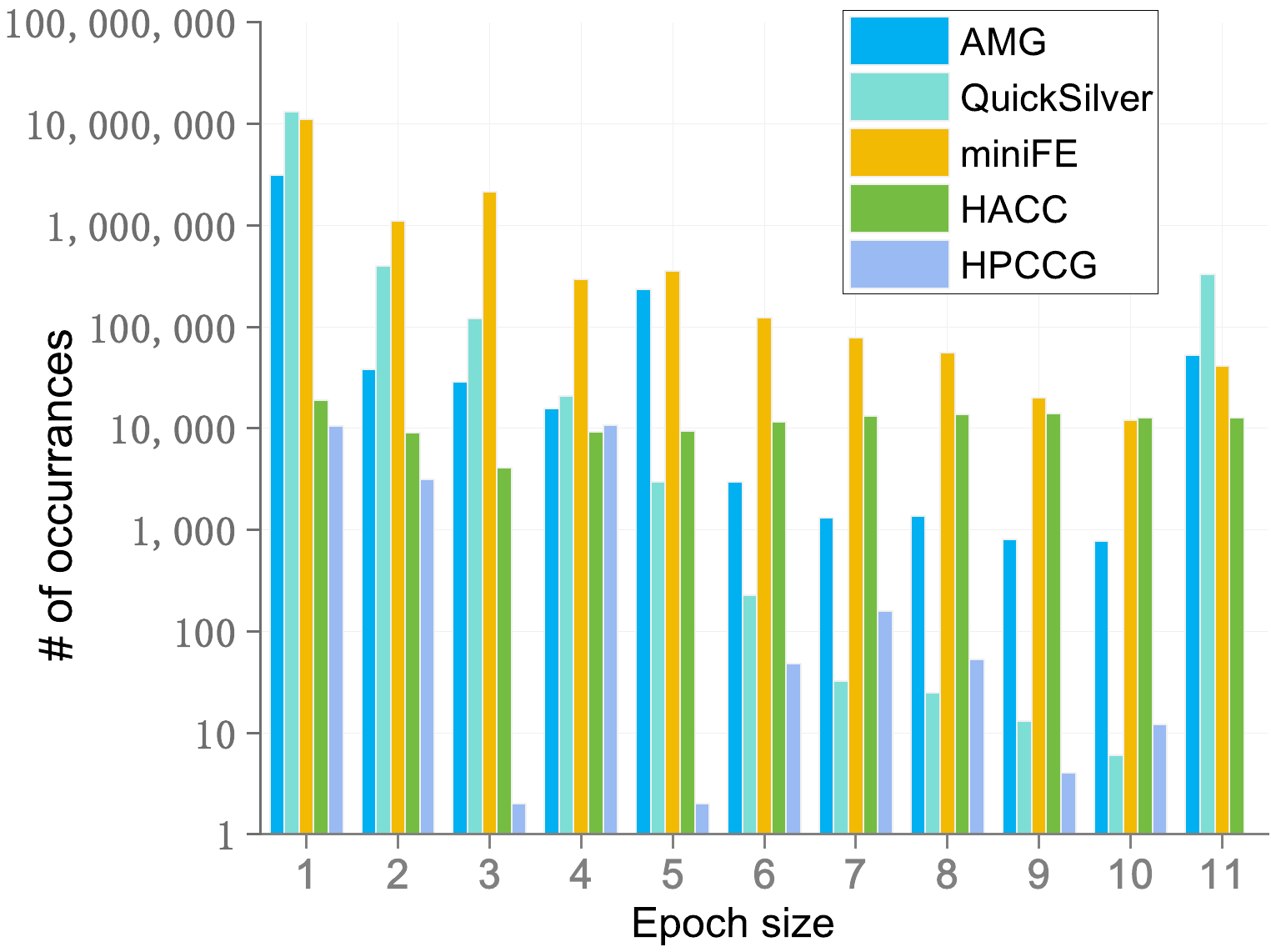}
 			\figcaption{The number of occurrences of each epoch size}
 			\label{fig:evaluation:profile}
		\end{minipage}
		\hspace{0.1cm}
		\begin{minipage}{5.5cm}
			\def\@captype{table}
			\centering
			\footnotesize
			\begin{tabular}{|c|c|c|c|c|}
				\hline
				\rowcolor{darkgray}   	& \multicolumn{2}{c|}{Record} & \multicolumn{2}{c|}{Replay} \\
				\cline{2-5}
				\rowcolor{darkgray}   	& DC & DE & DC & DE \\
				\hline
				\rowcolor{white}
				\cellcolor{lightgray} AMG& 0.97& 0.95& 3.32& 4.49\\
				\hline 
				\cellcolor{lightgray} QuickSilver & 1.05& 1.02& 1.93& 2.06\\
				\hline
				\cellcolor{lightgray} miniFE      & 1.11& 1.15& 2.87& 3.58\\
				\hline
				\cellcolor{lightgray} HACC& 1.2& 1.29& 4.01& 5.61\\    
				\hline
				\cellcolor{lightgray} HPCCG& 0.97& 0.90& 1.91& 3.37\\    
                \hline
			\end{tabular}
			\vspace{0.5cm}
    			\tblcaption{Factors of performance improvement of DC/DE recording over ST recording in 112 threads}
    			\label{tbl:evaluation:factor}
		\end{minipage}               
	\end{tabular}
\end{center}
\end{table*}

We evaluate ST, DC and DE recording in five non-deterministic applications, AMG, QuickSilver, miniFE, HACC and HPCCG.
AMG is a solver that uses parallel algebraic multigrid methods to tackle linear systems that stem from unstructured grid issues. QuickSilver represents applications solving a simplified dynamic Monte Carlo particle transport problem. The miniFE represents implicit finite-element applications. The Hardware Accelerated Cosmology Code (HACC) is a sophisticated framework that employs particle-mesh techniques to simulate the evolutionary processes of mass within the universe. HPCCG is a benchmark code for conjugate gradient computations within a 3D chimney-shaped domain, designed to run efficiently on any number of processors.

Figure~\ref{fig:evaluation:amg_omp} to~\ref{fig:evaluation:hpccg_omp} shows execution times in different number of threads and different record-and-replay approaches. In Figure~\ref{fig:evaluation:amg_omp}, it is evident that after the number of threads exceeds or equals 8, AMG exhibits a linear trend in ST's replay time. This is attributed to the prolonged replay time of ST when running 8 or more threads, exceeding half an hour. For a better view of the line graph, the execution time for 8 or more threads we set to 200 seconds.

In the ST model, the overhead of replay increases significantly as the number of threads increases, which is attributed to two key factors: intensive synchronization requirements and serialization constraints. Sequential consistency requires that all threads must replay operations exactly in the original order of execution, leading to frequent synchronization points in multithreaded environments. The cost of these synchronization operations accumulates rapidly as the number of threads increases. This serial bottleneck effect is especially pronounced for applications that rely on parallel efficiency, such as AMG.

In contrast, our DE technique is specifically designed to address these issues. By using distributed epochs, DE relaxes strict serialization requirements, allowing each thread to independently record and replay within its own epoch without enforcing global synchronization across the entire system. This mechanism significantly reduces unnecessary synchronization overhead while increasing the potential for parallel execution, even in the presence of a large number of threads. As a result, under the same experimental conditions, DC and DE exhibit significantly improved execution efficiency compared to ST in replay mode, as extensively discussed in Section~\ref{ssec:evaluation:benchmarks}.


Table~\ref{tbl:evaluation:factor} shows factors of performance improvement of DC and DE recording to ST recording 112 threads.
Replay runs by DC recording is faster than ones by ST recordings by factors of 3.32, 1.93, 2.87, 4.01 and 1.91 in AMG, QuickSilver, miniFE, HACC and HPCCG whereas DE recording further improves the performance by factors of 4.49, 2.06, 3.58, 5.61 and 3.37.

Figure~\ref{fig:evaluation:profile} shows the number of occurrences of each epoch size. 
Epoch size is the number of load or store instructions that belong to the same epoch.
For example, the sizes of epoch 0, 3, 5 and 6 in Table~\ref{tbl:sma:DE} are respectively 3, 2, 1 and 1.

Instructions with the same epoch can be parallelized, and if the epoch size is larger than 1, it means that there are multiple instructions that can be processed in parallel, which is the key of DE recording technique.

In fact, we can view DC records as a special case where each epoch is strictly limited to containing only one load or store instruction, lacking the flexibility of parallel execution. 
In contrast, DE technique, through its flexible distributed design,  effectively identifies and utilizes these parallel potentials.

In a deep analysis of several key applications, 10.6\%, 27.5\%, 85\%, and 57\% of epochs in AMG, miniFE, HACC, and HPCCG are epochs whose sizes are more than 1. However, the ratio in QuickSilver is only 4\% , indicating fewer opportunities for concurrent instructions. Therefore, the performance improvements in AMG, miniFE, HACC, and HPCCG by DE recording are more significant than the one in QuickSilver.

In conclusion, DE works better in applications where the proportion of epochs whose sizes greater than 1 is significant, which is important for reducing overhead.

\subsection{Case Study: Integration with ReMPI}
\label{ssec:evaluation:case_study}
As we described reproducibility challenges for debugging and testing in Section~\ref{sec:motivation}, we must record and replay both message passing and shared-memory access events for reproducing behavior of hybrid parallel applications. In these complex applications, a complex network of interactions between MPI processes and OpenMP threads is formed through MPI message passing and shared memory access. Specifically,  potential race conditions in MPI message passing and shared memory accesses make the communication sequence between processes and threads unpredictable. This uncertainty is not only reflected in the timing of data exchange, but also extends to the execution order of floating-point operations, which leads to inconsistent numerical results across executions.

For the recording and replaying of non-deterministic MPI+OpenMP applications, our carefully-determined ReOMP design enables hybrid use of ReOMP+ReMPI by simply preloading ReMPI library to maximize the usability.

If an application does not require \mona{MPI\_THREAD\_MULTIPLE} for the level of thread support, independent application of ReMPI and \toolname is enough for reproducing the application's behavior.
However, if the application does require \mona{MPI\_THREAD\_MULTIPLE}, different threads can potentially receive different messages
from run to run depending on which threads call MPI functions.
We can easily record and replay such non-deterministic MPI calls by 
instrumenting the \gateinout functions before and after MPI
receive~(\mona{MPI\_Recv}), wait~(\mona{MPI\_Wait} family),
test~(\mona{MPI\_Test} family) and probe~(\mona{MPI\_Probe/Iprobe}) functions. Our design choice easily enables such a composition. 

Finally, we apply ReMPI+\toolname to MPI+OpenMP applications HACC and HPCCG that we described in Section~\ref{ssec:motivation:challenge}.


Figure~\ref{fig:evaluation:mpi_omp_HACC} and Figure~\ref{fig:evaluation:mpi_omp_hpccg} show the execution times of HACC and HPCCG in
different number of threads. 
According to the topology of the HOKUSAI BigWaterfall2, we experimented with these applications with different combinations of nodes, processes, and threads with a fixed input data size (i.e., number of threads ranging from 24 to 4800). As can be seen from the figures, our approach is able to record and successfully replay the applications with small additional overhead.

In \toolname, each thread can record and replay its own record file independently, thus avoiding additional communication between threads throughout the process. This scalable feature of ReOMP remains unchanged even as the size of the application grows, the number of processes grows, and the number of spanning nodes grows, allowing each thread to focus on accessing its own record file without having to wait for access to a single, shared record file, as is the case with ST.

As long as the running application is scalable, then if we can use node-local storage or node-local memory space to store these record files instead of a shared file system as in this evaluation, then ReMPI+ReOMP will also become scalable as the number of processes increases. Using node-local storage or node-local memory space instead of a shared file system can greatly reduce I/O latency and improve data access speed, especially in massively parallel applications, where applying our approach, each thread can access its own record file quickly and independently for efficient replay.
\section{Related work}
An ability to record non-deterministic behaviors and deterministically replay
significantly facilitates debugging and testing non-deterministic applications.
Depending on the purpose, there are several levels of reproducibility.

Aftersight~\cite{VMM:2008}, ReVirt~\cite{ReVirt:2002},
\textsc{SCRIBE}~\cite{SCRIBE:2010} and Instant~Replay~\cite{InstantReplay:1987}
reproduce behaviors of an entire system by logging events at the level of a virtual machine or an operating system.
These approaches allow for diagnosing not only a single process but also
interaction among processes running on the same machine.
Although these
approaches give lower levels of reproducibility, they are too
expensive to debug and test just a single hybrid parallel application.

When reproducing behaviors of non-deterministic hybrid parallel applications for
debugging and testing, user-level record-and-replay is more useful
and lightweight. RecPlay~\cite{RecReplay:1999} and iDNA~\cite{iDNA:2006}
instrument and trace synchronization operations, i.e., lock and unlock
events for critical sections. However, these tools do not have an
ability to record and replay applications that have data races.
As we have seen in Section~\ref{ssec:evaluation:proxy}, applications can potentially have data races, especially benign data races,
leading to non-determinism. Recording and replaying all shared-memory
accesses are necessary for reproducing hybrid parallel applications.

There are several tools that have the ability to reproduce even data
racy programs. Totalview ReplayEngine~\cite{TVRE:2024} records the execution
history of how variables are updated.
Although this replay engine can record and replay data racy shared-memory accesses, it introduces
large performance and memory overhead to applications since it records everything, i.e., both
deterministic and non-deterministic execution history.
PinPlay~\cite{PinPlay:2010} records and replays data races by dynamically
instrumenting data racy shared-memory accesses detected by the FDR tool~\cite{FDR:2003}.
Unfortunately, dynamic instrumentation also causes large overhead to applications.

Chimera~\cite{Chimera:2012} adopts the most similar approach to our approach.
Chimera first applies a data-race detector and then instruments pairs of
potentially data racy instructions with \mutexlockunlock by using a
source-to-source compiler.
However, Chimera uses \emph{weak-lock}. 
Their weak-lock is a time-out lock which times out if not acquired in reasonable amount of time,
an event that can lead to incorrect replay.

By contrast, we proposed a novel DE recording.
The distributed recording and the use of epoch will be able to complement some of the existing shared memory and message-passing record-and-replay techniques with significant performance and scalability properties. Equally importantly, although there exist several record-and-replay tools, they are for either only message passing or shared memory. To the best of our knowledge, our ReMPI+\toolname integration is the first successful demonstration of scalably recording and replying OpenMP+MPI parallel applications using a novel and efficient DE recording technique.

Non-determinism event record-and-replay is also a valid technique used in distributed system fault tolerance. For instance, record-and-replay is key when a failed process is rolled back to assure consistent shared-memory access order to data restored from checkpoints. 
HyCoR~\cite{2021HyCoR}, proposed by Zhou and Tamir, uses a hybrid of checkpointing and replay to achieve fault tolerance. By periodically sending checkpoints and recording non-deterministic events, HyCoR minimizes client output delays and reduces recovery time upon a primary failure.
Checkpoint/Restart (C/R) techniques periodically save the state of an application during its runtime (checkpointing), allowing rapid recovery (restarting) from the latest checkpoint in case of failures. Depending on the scope and granularity of the checkpointing process, C/R techniques can be categorized into two main types: system-level C/R~\cite{BLCR,DMTCP,sato2012design,moody2010design} and application-level C/R~\cite{FTI,VeloC,guo2020match,georgakoudis2020reinit}. 
In future work, we will 1) investigate more non-deterministic OpenMP scheduling such as parallel task scheduling; 2) explore potentials in advancing C/R with our distributed record-and-replay technique.

\section{Conclusion}
Efficient shared memory record-and-replay is critical since shared memory record-and-replay
overhead dominates the overall tool overhead. 
To minimize the overhead, we proposed novel and
efficient distributed clock recording (DC recording) and distributed epoch recording (DE recording) and implemented them in \toolname.
And our techniques can also be applied to existing record-and-replay tools and can be orthogonalized to them.
Our evaluation shows that DC and DE recording can significantly outperform
the traditional approach, ST recording, in our synthetic OpenMP
benchmarks as well as five applications.
We also integrate \toolname with ReMPI and apply ReMPI+\toolname to two non-deterministic MPI+OpenMP applications. Our evaluation shows that ReMPI+\toolname can record
and successfully replay MPI+OpenMP applications with small overhead for recording and replaying at
scale.

	
\section*{Acknowledgment}
This work was performed under the auspices of the U.S. Department of Energy
by LLNL under contract DE-AC52-07NA27344 (LLNL-CONF-748571).


%


\bibliographystyle{IEEEtran}
\bibliography{reference}

@inproceedings{moody2010design,
  title={Design, modeling, and evaluation of a scalable multi-level checkpointing system},
  author={Moody, Adam and Bronevetsky, Greg and Mohror, Kathryn and De Supinski, Bronis R},
  booktitle={SC'10: Proceedings of the 2010 ACM/IEEE International Conference for High Performance Computing, Networking, Storage and Analysis},
  pages={1--11},
  year={2010},
  organization={IEEE}
}

@inproceedings{georgakoudis2020reinit,
  title={Reinit: Evaluating the performance of global-restart recovery methods for mpi fault tolerance},
  author={Georgakoudis, Giorgis and Guo, Luanzheng and Laguna, Ignacio},
  booktitle={International Conference on High Performance Computing},
  pages={536--554},
  year={2020},
  organization={Springer}
}

@inproceedings{guo2020match,
  title={Match: An mpi fault tolerance benchmark suite},
  author={Guo, Luanzheng and Georgakoudis, Giorgis and Parasyris, Konstantinos and Laguna, Ignacio and Li, Dong},
  booktitle={2020 IEEE International Symposium on Workload Characterization (IISWC)},
  pages={60--71},
  year={2020},
  organization={IEEE}
}

@inproceedings{sato2012design,
  title={Design and modeling of a non-blocking checkpointing system},
  author={Sato, Kento and Maruyama, Naoya and Mohror, Kathryn and Moody, Adam and Gamblin, Todd and de Supinski, Bronis R and Matsuoka, Satoshi},
  booktitle={SC'12: Proceedings of the International Conference on High Performance Computing, Networking, Storage and Analysis},
  pages={1--10},
  year={2012},
  organization={IEEE}
}

@INPROCEEDINGS{Drosinos:2004, 
author={N. Drosinos and N. Koziris}, 
booktitle={18th International Parallel and Distributed Processing Symposium, 2004. Proceedings.}, 
title={Performance comparison of pure MPI vs hybrid MPI-OpenMP parallelization models on SMP clusters}, 
year={2004}, 
volume={}, 
number={}, 
pages={15-}, 
doi={10.1109/IPDPS.2004.1302919}, 
ISSN={}, 
month={April},}

@article{Mininni:2011,
title = "A hybrid MPI–OpenMP scheme for scalable parallel pseudospectral computations for fluid turbulence",
journal = "Parallel Computing",
volume = "37",
number = "6",
pages = "316 - 326",
year = "2011",
issn = "0167-8191",
doi = "https://doi.org/10.1016/j.parco.2011.05.004",
url = "http://www.sciencedirect.com/science/article/pii/S0167819111000512",
author = "Pablo D. Mininni and Duane Rosenberg and Raghu Reddy and Annick Pouquet"
}

@INPROCEEDINGS{Wolf:2003, 
author={F. Wolf and B. Mohr}, 
booktitle={Eleventh Euromicro Conference on Parallel, Distributed and Network-Based Processing, 2003. Proceedings.}, 
title={Automatic performance analysis of hybrid MPI/OpenMP applications}, 
year={2003}, 
volume={}, 
number={}, 
pages={13-22}, 
doi={10.1109/EMPDP.2003.1183560}, 
ISSN={1066-6192}, 
month={Feb},}

@article{Yang:2011,
author = {Yang, Chao-Tung and Huang, Chih-Lin and Lin, Cheng-Fang},
year = {2011},
month = {01},
pages = {266-269},
title = {Hybrid CUDA, OpenMP, and MPI parallel programming on multicore GPU clusters},
volume = {182},
booktitle = {Computer Physics Communications}
}

@INPROCEEDINGS{Rabenseifner:2009, 
author={R. Rabenseifner and G. Hager and G. Jost}, 
booktitle={2009 17th Euromicro International Conference on Parallel, Distributed and Network-based Processing}, 
title={Hybrid MPI/OpenMP Parallel Programming on Clusters of Multi-Core SMP Nodes}, 
year={2009}, 
volume={}, 
number={}, 
pages={427-436}, 
doi={10.1109/PDP.2009.43}, 
ISSN={1066-6192}, 
month={Feb},}

@inproceedings{Boehm:2011,
 author = {Boehm, Hans-J.},
 title = {How to Miscompile Programs with "Benign" Data Races},
 booktitle = {Proceedings of the 3rd USENIX Conference on Hot Topic in Parallelism},
 series = {HotPar'11},
 year = {2011},
 location = {Berkeley, CA},
 pages = {3--3},
 numpages = {1},
 acmid = {2001255},
 publisher = {USENIX Association},
 address = {Berkeley, CA, USA},
}

@misc{LLVMOpenMP,
	author = {clang/LLVM},
	title = {{LLVM OpenMP Runtime Library}},
	url = {https://openmp.llvm.org/Reference.pdf},
	year = {As of May 8, 2024}
}

@INPROCEEDINGS{Richards:2017,
author={D. F. Richards and R. C. Bleile and P. S. Brantley and S. A. Dawson and M. S. McKinley and M. J. O’Brien}, 
booktitle={2017 IEEE International Conference on Cluster Computing (CLUSTER)}, 
title={Quicksilver: A Proxy App for the Monte Carlo Transport Code Mercury}, 
year={2017}, 
volume={}, 
number={}, 
pages={866-873}, 
doi={10.1109/CLUSTER.2017.121}, 
ISSN={}, 
month={Sept},}

@misc{AMG,
    url = {https://github.com/LLNL/AMG},
    title = {{LLNL/AMG}},
    year = {As of May 8, 2024},
}

@misc{QuickSilver,
	author = {D. F. Richards},
    url = {https://github.com/LLNL/Quicksilver},
    title = {{LLNL/QuickSilver}},
    year = {As of May 8, 2024},
}

@misc{ReMPI,
	author = {Kento Sato},
    url = {https://github.com/PRUNERS/ReMPI},
    title = {{ReMPI}},
    year = {As of May 8, 2024},
}

@incollection{Kranzlmuller:2001,
year={2001},
isbn={978-3-540-42609-7},
booktitle={Recent Advances in Parallel Virtual Machine and Message Passing Interface},
volume={2131},
series={Lecture Notes in Computer Science},
doi={10.1007/3-540-45417-9_28},
title={{An Integrated Record \& Replay Mechanism for Nondeterministic Message Passing Programs}},
url={http://dx.doi.org/10.1007/3-540-45417-9_28},
publisher={Springer Berlin Heidelberg},
author={Kranzlm\"uller, Dieter and Schaubschl\"ager, Christian and Volkert, Jens},
pages={192-200},
language={English}
}

@inproceedings{CDC:2015,
 author = {Sato, Kento and Ahn, Dong H. and Laguna, Ignacio and Lee, Gregory L. and Schulz, Martin},
 title = {Clock Delta Compression for Scalable Order-replay of Non-deterministic Parallel Applications},
 booktitle = {Proceedings of the International Conference for High Performance Computing, Networking, Storage and Analysis},
 series = {SC '15},
 year = {2015},
 isbn = {978-1-4503-3723-6},
 location = {Austin, Texas},
 pages = {62:1--62:12},
 articleno = {62},
 numpages = {12},
 url = {http://doi.acm.org/10.1145/2807591.2807642},
 doi = {10.1145/2807591.2807642},
 acmid = {2807642},
 publisher = {ACM},
 address = {New York, NY, USA},
}

@inproceedings{MPIWiz:Xue:2009,
 author = {Xue, Ruini and Liu, Xuezheng and Wu, Ming and Guo, Zhenyu and Chen, Wenguang and Zheng, Weimin and Zhang, Zheng and Voelker, Geoffrey},
 title = {{MPIWiz: Subgroup Reproducible Replay of Mpi Applications}},
 booktitle = {Proceedings of the 14th ACM SIGPLAN Symposium on Principles and Practice of Parallel Programming},
 series = {PPoPP '09},
 year = {2009},
 isbn = {978-1-60558-397-6},
 location = {Raleigh, NC, USA},
 pages = {251--260},
 numpages = {10},
 url = {http://doi.acm.org/10.1145/1504176.1504213},
 doi = {10.1145/1504176.1504213},
 acmid = {1504213},
 publisher = {ACM},
 address = {New York, NY, USA},
}

@inproceedings{NINJA:2017,
 author = {Sato, Kento and Ahn, Dong H. and Laguna, Ignacio and Lee, Gregory L. and Schulz, Martin and Chambreau, Christopher M.},
 title = {Noise Injection Techniques to Expose Subtle and Unintended Message Races},
 booktitle = {Proceedings of the 22Nd ACM SIGPLAN Symposium on Principles and Practice of Parallel Programming},
 series = {PPoPP '17},
 year = {2017},
 isbn = {978-1-4503-4493-7},
 location = {Austin, Texas, USA},
 pages = {89--101},
 numpages = {13},
 url = {http://doi.acm.org/10.1145/3018743.3018767},
 doi = {10.1145/3018743.3018767},
 acmid = {3018767},
 publisher = {ACM},
 address = {New York, NY, USA},
}

@inproceedings{DOpenMP:2011,
 author = {Aviram, Amittai and Ford, Bryan},
 title = {Deterministic OpenMP for Race-free Parallelism},
 booktitle = {Proceedings of the 3rd USENIX Conference on Hot Topic in Parallelism},
 series = {HotPar'11},
 year = {2011},
 location = {Berkeley, CA},
 pages = {4--4},
 numpages = {1},
 url = {http://dl.acm.org/citation.cfm?id=2001252.2001256},
 acmid = {2001256},
 publisher = {USENIX Association},
 address = {Berkeley, CA, USA},
}

@INPROCEEDINGS{Archer:2016, 
author={S. Atzeni and G. Gopalakrishnan and Z. Rakamaric and D. H. Ahn and I. Laguna and M. Schulz and G. L. Lee and J. Protze and M. S. Müller}, 
booktitle={2016 IEEE International Parallel and Distributed Processing Symposium (IPDPS)}, 
title={ARCHER: Effectively Spotting Data Races in Large OpenMP Applications}, 
year={2016}, 
volume={}, 
number={}, 
pages={53-62}, 
doi={10.1109/IPDPS.2016.68}, 
ISSN={1530-2075}, 
month={May},}

@misc{OpenMP,
	author = {{OpenMP ARB (Architecture Review Boards)}},
    url = {http://www.openmp.org/},
    title = {{OpenMP}},
    year = {As of May 8, 2024}
}

@misc{MPI,
	author = {{MPI working group}},
    url = {http://mpi-forum.org/},
    title = {{MPI Forum}},
    year = {As of May 8, 2024}
}

@misc{top500,
  author = {{Erich Strohmaier and Jack Dongarra and Horst Simon and Martin Meuer}},
  title = {{Top 500}},
  url = {https://www.top500.org/},
  year = {As of May 8, 2024},  
}

@misc{HBW2,
  author = {{RIKEN}},
  title = {{Supercomputer System HBW2 TOP | ISC, RIKEN}},
  url = {https://i.riken.jp/en/supercom/},
  year = {As of May 8, 2024},  
}

@misc{sierra,
  author = {{Lawrence Livermore National Laboratory}},
  title = {{CORAL/Sierra}},
  url = {https://asc.llnl.gov/coral-info},
  year = {As of May 8, 2024},  
}

@misc{summit,
  author = {{Oak Ridge National Laboratory}},
  title = {{Summit - Oak Ridge Leadership Facility}},
  url = {https://www.olcf.ornl.gov/olcf-resources/compute-systems/summit/}, 
  year = {As of May 8, 2024},   
}

@misc{CORAL_benchmarks,
	title = {{CORAL Benchmarks}},
    howpublished = {https://codesign.llnl.gov/mcb.php}
}

@misc{HPCCG,
    author = {Michael A. Heroux, Sandia National Laboratories},
    url = {https://github.com/BenWibking/hacc-coral/tree/master},
    title = {{Mantevo/HPCCG}},
    year = {2017},
}

@misc{HACC,
    url = {https://github.com/BenWibking/hacc-coral/tree/master},
    title = {{BenWibking/hacc-coral}},
}

@inproceedings{iDNA:2006,
 author = {Bhansali, Sanjay and Chen, Wen-Ke and de Jong, Stuart and Edwards, Andrew and Murray, Ron and Drini\'{c}, Milenko and Miho\v{c}ka, Darek and Chau, Joe},
 title = {Framework for Instruction-level Tracing and Analysis of Program Executions},
 booktitle = {Proceedings of the 2Nd International Conference on Virtual Execution Environments},
 series = {VEE '06},
 year = {2006},
 isbn = {1-59593-332-8},
 location = {Ottawa, Ontario, Canada},
 pages = {154--163},
 numpages = {10},
 url = {http://doi.acm.org/10.1145/1134760.1220164},
 doi = {10.1145/1134760.1220164},
 acmid = {1220164},
 publisher = {ACM},
 address = {New York, NY, USA},
}

@inproceedings{Chimera:2012,
 author = {Lee, Dongyoon and Chen, Peter M. and Flinn, Jason and Narayanasamy, Satish},
 title = {Chimera: Hybrid Program Analysis for Determinism},
 booktitle = {Proceedings of the 33rd ACM SIGPLAN Conference on Programming Language Design and Implementation},
 series = {PLDI '12},
 year = {2012},
 isbn = {978-1-4503-1205-9},
 location = {Beijing, China},
 pages = {463--474},
 numpages = {12},
 url = {http://doi.acm.org/10.1145/2254064.2254119},
 doi = {10.1145/2254064.2254119},
 acmid = {2254119},
 publisher = {ACM},
 address = {New York, NY, USA},
}

@article{RecReplay:1999,
 author = {Ronsse, Michiel and De Bosschere, Koen},
 title = {RecPlay: A Fully Integrated Practical Record/Replay System},
 journal = {ACM Trans. Comput. Syst.},
 issue_date = {May 1999},
 volume = {17},
 number = {2},
 month = may,
 year = {1999},
 issn = {0734-2071},
 pages = {133--152},
 numpages = {20},
 url = {http://doi.acm.org/10.1145/312203.312214},
 doi = {10.1145/312203.312214},
 acmid = {312214},
 publisher = {ACM},
 address = {New York, NY, USA},
}

@inproceedings{PinPlay:2010,
 author = {Patil, Harish and Pereira, Cristiano and Stallcup, Mack and Lueck, Gregory and Cownie, James},
 title = {PinPlay: A Framework for Deterministic Replay and Reproducible Analysis of Parallel Programs},
 booktitle = {Proceedings of the 8th Annual IEEE/ACM International Symposium on Code Generation and Optimization},
 series = {CGO '10},
 year = {2010},
 isbn = {978-1-60558-635-9},
 location = {Toronto, Ontario, Canada},
 pages = {2--11},
 numpages = {10},
 url = {http://doi.acm.org/10.1145/1772954.1772958},
 doi = {10.1145/1772954.1772958},
 acmid = {1772958},
 publisher = {ACM},
 address = {New York, NY, USA},
}

@inproceedings{FDR:2003,
 author = {Xu, Min and Bodik, Rastislav and Hill, Mark D.},
 title = {A "Flight Data Recorder" for Enabling Full-system Multiprocessor Deterministic Replay},
 booktitle = {Proceedings of the 30th Annual International Symposium on Computer Architecture},
 series = {ISCA '03},
 year = {2003},
 isbn = {0-7695-1945-8},
 location = {San Diego, California},
 pages = {122--135},
 numpages = {14},
 url = {http://doi.acm.org/10.1145/859618.859633},
 doi = {10.1145/859618.859633},
 acmid = {859633},
 publisher = {ACM},
 address = {New York, NY, USA},
}

@misc{TVRE:2024,
    url = {http://www.roguewave.com/products-services/features/reverse-debugging},
    title = {{Reverse debugging with ReplayEngine. Rogue Wave Software}},
    year = {As of May 8, 2024}
}

@misc{TV:2024,
    url = {http://www.roguewave.com/products-services/totalview},
    title = {{TotalView for HPC. Rogue Wave Software}},
    year={{As of May 8, 2024}}
}

@misc{DDT:2024,
    url = {https://www.allinea.com/products/ddt},
    title = {{Allinea DDT (Distributed Debugging Tool). Allinea Software (Now Part of ARM)}},
    year={{As of May 8, 2024}}
}

@article{InstantReplay:1987,
 author = {LeBlanc, T. J. and Mellor-Crummey, J. M.},
 title = {Debugging Parallel Programs with Instant Replay},
 journal = {IEEE Trans. Comput.},
 issue_date = {April 1987},
 volume = {36},
 number = {4},
 month = apr,
 year = {1987},
 issn = {0018-9340},
 pages = {471--482},
 numpages = {12},
 url = {http://dx.doi.org/10.1109/TC.1987.1676929},
 doi = {10.1109/TC.1987.1676929},
 acmid = {32396},
 publisher = {IEEE Computer Society},
 address = {Washington, DC, USA},
}

@inproceedings{SCRIBE:2010,
 author = {Laadan, Oren and Viennot, Nicolas and Nieh, Jason},
 title = {Transparent, Lightweight Application Execution Replay on Commodity Multiprocessor Operating Systems},
 booktitle = {Proceedings of the ACM SIGMETRICS International Conference on Measurement and Modeling of Computer Systems},
 series = {SIGMETRICS '10},
 year = {2010},
 isbn = {978-1-4503-0038-4},
 location = {New York, New York, USA},
 pages = {155--166},
 numpages = {12},
 url = {http://doi.acm.org/10.1145/1811039.1811057},
 doi = {10.1145/1811039.1811057},
 acmid = {1811057},
 publisher = {ACM},
 address = {New York, NY, USA},
}

@article{ReVirt:2002,
 author = {Dunlap, George W. and King, Samuel T. and Cinar, Sukru and Basrai, Murtaza A. and Chen, Peter M.},
 title = {ReVirt: Enabling Intrusion Analysis Through Virtual-machine Logging and Replay},
 journal = {SIGOPS Oper. Syst. Rev.},
 issue_date = {Winter 2002},
 volume = {36},
 number = {SI},
 month = dec,
 year = {2002},
 issn = {0163-5980},
 pages = {211--224},
 numpages = {14},
 url = {http://doi.acm.org/10.1145/844128.844148},
 doi = {10.1145/844128.844148},
 acmid = {844148},
 publisher = {ACM},
 address = {New York, NY, USA},
}

@inproceedings{VMM:2008,
 author = {Chow, Jim and Garfinkel, Tal and Chen, Peter M.},
 title = {Decoupling Dynamic Program Analysis from Execution in Virtual Environments},
 booktitle = {USENIX 2008 Annual Technical Conference},
 series = {ATC'08},
 year = {2008},
 location = {Boston, Massachusetts},
 pages = {1--14},
 numpages = {14},
 url = {http://dl.acm.org/citation.cfm?id=1404014.1404015},
 acmid = {1404015},
 publisher = {USENIX Association},
 address = {Berkeley, CA, USA},
}

@inproceedings{Tsan,
author = {Serebryany, Konstantin and Iskhodzhanov, Timur},
title = {ThreadSanitizer: data race detection in practice},
year = {2009},
isbn = {9781605587936},
publisher = {Association for Computing Machinery},
address = {New York, NY, USA},
url = {https://doi.org/10.1145/1791194.1791203},
doi = {10.1145/1791194.1791203},
booktitle = {Proceedings of the Workshop on Binary Instrumentation and Applications},
pages = {62–71},
numpages = {10},
location = {New York, New York, USA},
series = {WBIA '09}
}

@inproceedings{BLCR,
  title={Berkeley lab checkpoint/restart (blcr) for linux clusters},
  author={Hargrove, Paul H and Duell, Jason C},
  booktitle={Journal of Physics: Conference Series},
  volume={46},
  number={1},
  pages={067},
  year={2006},
  organization={IOP Publishing}
}

@inproceedings{DMTCP,
  title={DMTCP: Transparent checkpointing for cluster computations and the desktop},
  author={Ansel, Jason and Arya, Kapil and Cooperman, Gene},
  booktitle={2009 IEEE international symposium on parallel \& distributed processing},
  pages={1--12},
  year={2009},
  organization={IEEE}
}

@inproceedings{FTI,
  title={FTI: High performance fault tolerance interface for hybrid systems},
  author={Bautista-Gomez, Leonardo and Tsuboi, Seiji and Komatitsch, Dimitri and Cappello, Franck and Maruyama, Naoya and Matsuoka, Satoshi},
  booktitle={Proceedings of 2011 international conference for high performance computing, networking, storage and analysis},
  pages={1--32},
  year={2011}
}

@inproceedings{VeloC,
  title={Veloc: Towards high performance adaptive asynchronous checkpointing at large scale},
  author={Nicolae, Bogdan and Moody, Adam and Gonsiorowski, Elsa and Mohror, Kathryn and Cappello, Franck},
  booktitle={2019 IEEE International Parallel and Distributed Processing Symposium (IPDPS)},
  pages={911--920},
  year={2019},
  organization={IEEE}
}

@article{2021HyCoR,
  author = {Diyu Zhou and Yuval Tamir},
  title = {HyCoR: Fault-Tolerant Replicated Containers Based on Checkpoint and Replay},
  year = {2021},
  archivePrefix = {arXiv},
  eprint = {2101.09584},
  primaryClass = {cs.DC},
  doi = {10.48550/arXiv.2101.09584},
  url = {https://arxiv.org/abs/2101.09584}
}

%
\end{document}